\documentclass[twocolumn,showpacs,superscriptaddress,amssymb,hyperlink]{revtex4-2}
\usepackage{amsmath,amssymb,wasysym,ytableau}
\usepackage{graphicx}
\usepackage{bm}
\usepackage{braket, slashed}
\usepackage{xcolor}
\usepackage{verbatim}
\usepackage{float}
\usepackage{caption,subcaption}
\usepackage[T1]{fontenc}
\usepackage[colorlinks=true ,urlcolor=blue,citecolor=blue,urlbordercolor={0 1 1}]{hyperref}
\usepackage{feynmp-auto}
\usepackage{ulem}

\renewcommand{\ol}{\overline}
\newcommand{\del}{\partial}
\renewcommand{\vec}[1]{\mathbf{#1}}
\newcommand{\sla}[1]{\slashed{#1}}
\renewcommand{\Re}{\operatorname{Re}}
\renewcommand{\Im}{\operatorname{Im}}

\begin{document}
\title{When Wannier centers jump: Critical points between atomic insulating phases}
\author{Yunchao Zhang} 
\affiliation{Department of Physics, Massachusetts Institute of Technology, Cambridge MA 02139-4307,  USA}

\author{T. Senthil}
\affiliation{Department of Physics, Massachusetts Institute of Technology, Cambridge MA 02139-4307,  USA}

\date{\today} 

\begin{abstract}
We study a class of quantum phase transitions between featureless bosonic atomic insulators in $(2+1)$ dimensions, where each phase exhibits neither topological order nor protected edge modes. Despite their lack of topology, these insulators may be ``obstructed'' in the sense that their Wannier centers are not pinned to the physical atomic sites. 
These insulators represent distinct phases, as no symmetry-preserving adiabatic path connects them.
Surprisingly, we find that for certain lattices,
the critical point between these insulators can host a conformally invariant state described by quantum electrodynamics in $(2+1)$ dimensions (QED$_3$).
The emergent electrodynamics at the critical point can be stabilized if the embedding of the microscopic lattice symmetries suppresses the proliferation of monopoles, suggesting that even transitions between trivial phases can harbor rich and unexpected physics. 
We analyze the mechanism behind this phenomenon, discuss its stability against perturbations, and explore the embedding of lattice symmetries into the continuum through anomaly matching.
In all the models we analyze, we confirm that the QED$_3$ is indeed emergeable, in the sense that it is realizable from a local lattice Hamiltonian.
\end{abstract}

\maketitle
\section{Introduction}
Atomic insulators are typically regarded as the most mundane of quantum phases; 
such systems are gapped to all excitations, and they lack topological invariants, protected edge modes, or any form of long-range entanglement. 
Furthermore, their ground states, by definition, can be adiabatically deformed into trivial product states without closing the bulk gap. 
However, a large body of work has revealed a subtle distinction within this class; while some atomic insulators are trivial in the strictest sense (admitting symmetric, exponentially localized Wannier functions centered at lattice sites), others are ``obstructed'' \cite{bradlyn_nature_2017,po_prl_2018,khalaf_prr_2021,cano_annrevcm_2021,xu_prb_2024}. 
The obstructed atomic insulators are distinguished from the trivial class in that their Wannier centers lie away from the physical lattice sites.

Classes of obstructed atomic insulators (OAIs) have been studied as examples of higher-order topological insulators (HOTIs) \cite{bbh_prb_2017,bbh_sci_2017,schindler_sci_2018}, where their non-triviality manifests when the system is terminated with boundaries.
Unlike symmetry protected topological phases, which host anomalous gapless
surface states, OAIs host surface states of lower dimensionality, such as gapless corner modes on a $(2+1)$ dimensional lattice.

However, even in the bulk, OAIs are distinct from trivial insulators.
While obstructed atomic insulators preserve all symmetries of the lattice space group, they admit no smooth deformation to an on-site atomic limit without breaking symmetries. 

We emphasize that the notion of which phase is trivial and which is obstructed is only a relative concept.
This is because one must make an arbitrary
choice, such as where to place the lattice sites or how to cut the boundary in a finite system, in order to designate a phase as the trivial phase \cite{po_nature_2017,bradlyn_nature_2017}.
One
example of this is in the Su–Schrieffer–Heeger
(SSH) chain \cite{su_prl_1979}, in which the two phases have different polarization values that can be distinguished only relative to each other
\footnote{This relative nontriviality is sometimes presented as an absolute one, but any such distinction requires fixing a convention for the unit cell or equivalently, fixing the locations of a background of positive ions.
Mathematically, these relative topological distinctions are captured by
the mathematical structure of a torsor, not a group or cohomology ring as is usual for the study of topological phases.}.

\begin{figure}[t]
\captionsetup{justification=raggedright}
{\includegraphics[width=\columnwidth]{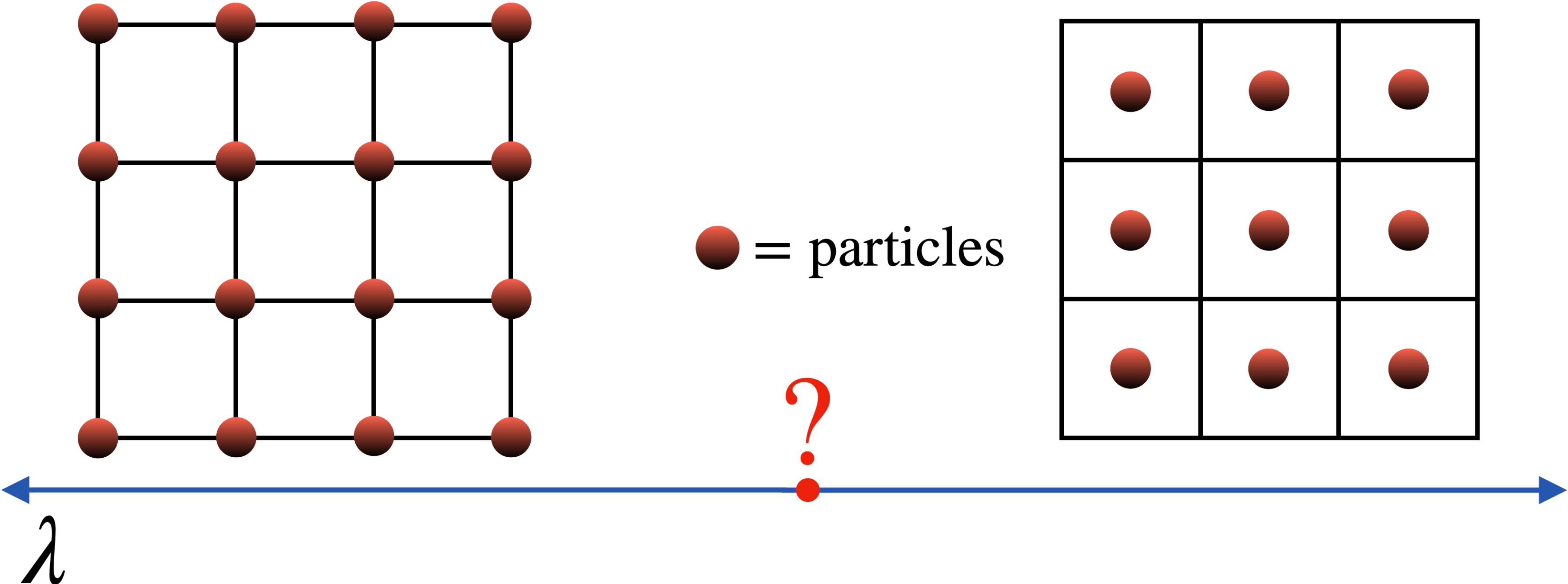}}
\caption{
Two possible atomic insulating phases on the square lattice, tuned by a parameter $\lambda$. In the language of crystallographic space groups, the Wannier centers of the particles move from the $1a$ to the $1d$ Wyckoff
positions as $\lambda$ is tuned across the phase transition.}
\label{fig:teaser}
\end{figure}

This raises a natural question: What can happen at a quantum phase transition between two such ``trivial'' insulating phases?
To this end, we will consider critical points between atomic insulators in $(2+1)d$.
A simple example can be realized on the square lattice \cite{bbh_prb_2017,bbh_sci_2017}
with multiple orbitals per site, in which one can tune a phase transition between two insulating states; one phase with localized atomic orbitals that sit on the atomic sites and another phase where the orbitals are centered at each plaquette, as shown in Fig.~\ref{fig:teaser}. 
In this case, at a filling of $n\ne 0\pmod{4}$ particles per unit cell, there is no $C_4$ preserving deformation between the two phases.
Therefore, there must be a transition between the two insulating phases.

Despite taking place between two trivial phases, there is no order parameter description of the phase transition.
While two phases with the same microscopic symmetry can, in principle, be separated by a Landau-Ginzburg transition with an emergent scalar order parameter (as in the liquid-gas critical point, where the coarse-grained density is described by a single Ising order parameter), the exotic critical points we explore are not in the universality class of any Landau-Ginzburg theory.
Instead, the closing of the single particle gap leads to a metallic critical point described by an interacting gauge theory.
In fact, examples such as deconfined quantum critical points \cite{wang_prx_2017,senthil2023deconfined}
have illustrated that even transitions between conventional ordered phases can be beyond the Landau paradigm \cite{senthil_sci_2004,senthil_prb_2004}.

In this paper, we show that a similar phenomenon occurs between featureless insulators, in which the insulating phases neither break symmetry nor carry topological order.

We observe that this already is realized for the quantum critical point between the two phases of the one-dimensional SSH chain in the presence of interactions.
For both the bosonic and fermionic SSH chains, the quantum critical point realizes a Luttinger liquid, without quasiparticles.

Going to higher dimension, at the transition between two atomic insulators of \textit{bosons} in $(2+1)$ dimensions, we find critical points described by $N_f=4$ QED$_3$, a theory of $4$ massless Dirac fermions coupled to an emergent $U(1)$ gauge field.
This theory has previously appeared as an effective description of quantum phases of matter such as algebraic spin liquids \cite{rantner_prl_2001,hermele_prb_2004,hastings_prb_2000,hermele_prb_2005,hermele_prb_2005_erratum,ran_prl_2007,hermele_prb_2008}, which can arise as ground states of frustrated magnets. 
QED$_3$ can also arise as an ``unnecessary'' quantum critical point within lattice magnets \cite{zhang_scipost_2025,zhang2025unnecessaryquantumcriticalitysu3}.
Moreover, many magnetically ordered states can be obtained by tuning away from a lattice QED$_3$ critical point \cite{song_nat_2019}.
From analytic and numerical calculations \cite{Appelquist_prl_1988,chesterjhep2016,li_2022_jhep,poland_prd_2022,he_scipost_2022,li2022plb}, QED$_3$ is believed to be a conformal field theory (CFT) in the IR.
Therefore, in contrast to the theory of deconfined criticality \cite{senthil2023deconfined}, in which later work showed that the putative field theory of the transition was weakly first order and discontinuous \cite{nahum_prx_2015,ma_prb_2020}, the existing evidence for the conformality of $N_f=4$ QED$_3$ is very well established.

The stability of the QED$_3$ critical point and consequently, the exact nature of the phase transition, is chiefly governed by how the microscopic (UV) lattice symmetries embed into the emergent continuum (IR) symmetries of the quantum critical theory \cite{song_prx_2020,song_nat_2019,zhang_scipost_2025}. 
At the lattice level, space group symmetries such as discrete translations, rotations, and discrete symmetries such as time reversal impose stringent conditions on which operators can be realized in the low energy effective theory.
Even independent of the microscopic theory, the embedding of UV symmetries into the IR theory is subject to constraints from the  't Hooft anomalies of the UV theory, which arise from the Lieb-Schultz-Mattis-Oshikawa-Hastings (LSMOH) theorems \cite{lieb_ann_1961,oshikawa_prl_2000,hastings_2004_prb}.

If the symmetry embedding in the IR permits no relevant operators other than the single one needed to tune to the critical point, then QED$_3$ can emerge as a stable fixed point that describes a continuous transition between the two insulating phases.
However, if lattice symmetries are insufficient to protect the critical theory and a relevant monopole perturbation is allowed, then the QED$_3$ theory will confine, and the phase transition will generically become first-order. 

The rest of this paper is organized as follows.
As a warm-up to our study of atomic insulator transitions, we begin in Sec.~\ref{sec:ssh} by reviewing the paradigmatic Su-Schrieffer-Heeger (SSH) model \cite{su_prl_1979}, both in its fermionic and bosonic forms. 
This simple one-dimensional model exemplifies how obstructed atomic insulators, though topologically trivial in the absence of symmetry protection, can host distinct phases separated by a critical point and illustrates the key features that will be present in the two-dimensional models. 
In the fermionic case, the transition involves a Dirac fermion gaining mass, while the bosonic version, analyzed via both exact methods and a parton construction, introduces an emergent gauge field relative to the fermionic case.
In both cases, the phase transition can be described by Luttinger liquid theory.

In Sec.~\ref{sec:bbh}, we generalize the SSH model to two dimensions by analyzing models of atomic insulators on various lattices, such as the Benalcazar-Bernevig-Hughes (BBH) model \cite{bbh_prb_2017,bbh_sci_2017} on the square lattice. 
Like the SSH model, each fermionic atomic insulator features Dirac band touchings at the transition. 
For the bosonic analog, we employ a parton construction that fractionalizes the boson into two fermions coupled to an emergent gauge field, yielding a critical theory described by $N_f=4$ QED$_3$.
However, the intrinsic bipartiteness of the square and honeycomb lattices introduces a single symmetry-allowed monopole operator that destabilizes the fixed point.
Together with a symmetry-allowed mass that tunes the transition, this drives the theory toward confinement and produces a highly multicritical point; we argue the transition is generically first-order.
Key to our argument is the observation made in \cite{song_prx_2020} that on a bipartite lattice, $N_f=4$ QED$_3$ admits a deformation to $N_f=2$ QCD$_3$, from which one can argue there is always at least one monopole operator that is a singlet under all UV symmetries.

In contrast, the $C_3$ symmetric breathing kagome lattice is tripartite, and we demonstrate that monopoles are symmetry-disallowed,
making QED$_3$ a natural candidate universality class for a direct continuous transition.
This highlights the strong dependence of criticality in trivial insulator transitions on the underlying (UV) lattice structure.

In Sec.~\ref{sec:lsm_matching}, we employ anomaly matching arguments to support our analysis of the critical theories. By matching the LSMOH anomalies for lattice bosons in the IR \cite{zou_prx_2021,ye_scipost_2022}, we confirm that all discussed critical theories, even the unstable ones, are realizable by local lattice Hamiltonians. This establishes a rigorous link between the microscopic theory and emergent QED$_3$ critical point.
We conclude in Sec.~\ref{sec:outlook} with broader implications and comments on extensions to three dimensional obstructed atomic insulators.

\section{A warm-up in one dimension: The SSH chain}
\label{sec:ssh}
\subsection{Fermionic}
\begin{figure}[t]
\captionsetup{justification=raggedright}
{\includegraphics[width=\columnwidth]{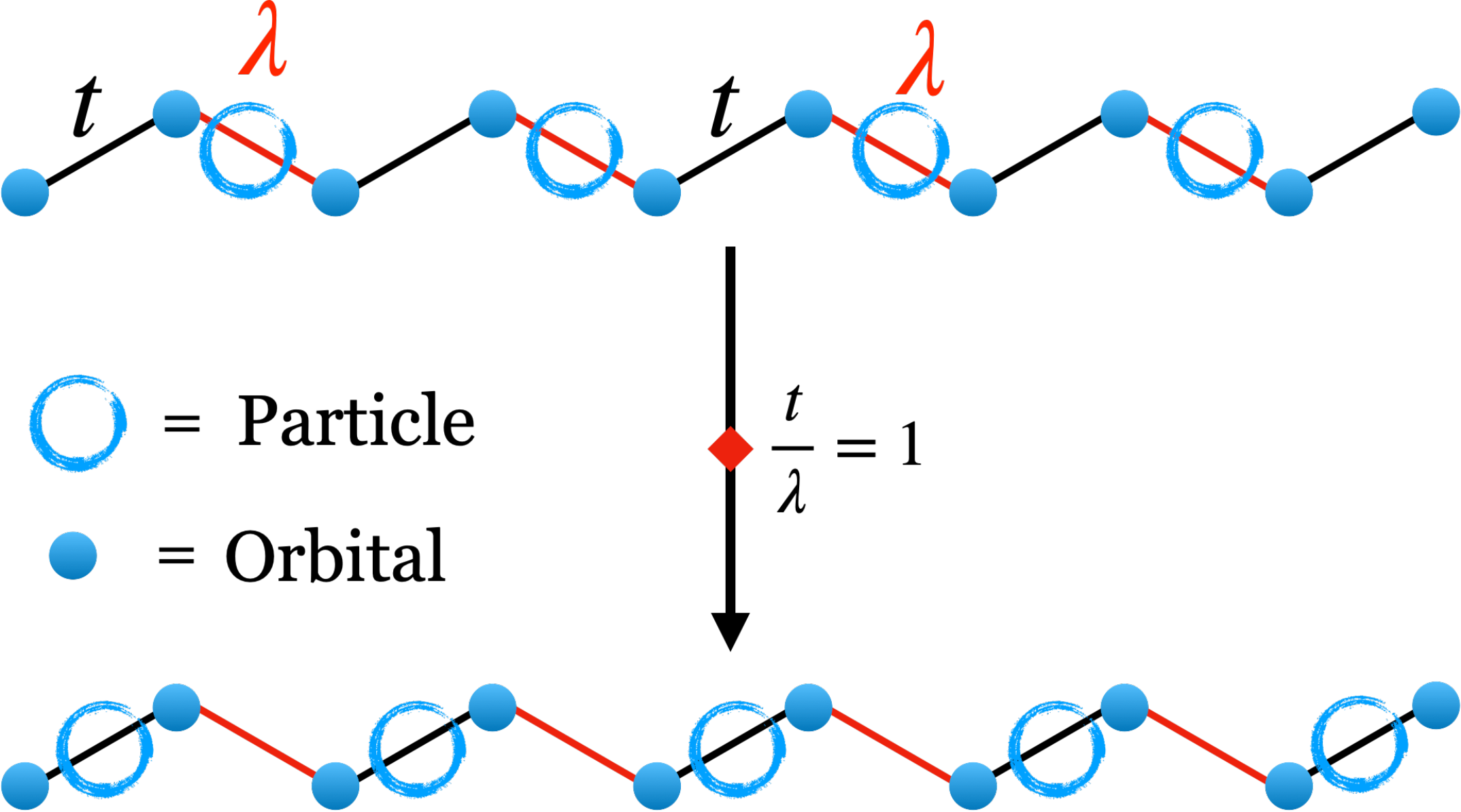}}
\caption{
An illustration of the two phases of the SSH model as the relative strength of the dimerized hoppings is tuned.}
\label{fig:ssh_fig}
\end{figure}

A minimal model to illustrate the general case is the SSH model \cite{su_prl_1979} in one dimension, which describes a chain with alternating bonds of two types between atoms, as in polyacetylene.
The fermionic SSH chain has a tight binding Hamiltonian 
\begin{equation}
    H_{SSH} = \sum_{n=1}^{N} \left( t c_{A,n}^\dagger c_{B,n} + \lambda c_{B,n}^\dagger c_{A,n+1} \right) + h.c.,
\end{equation}
which describes electrons hopping on a dimerized chain (with sublattice sites $A$ and $B$) at half filling.
Equivalently, one can also interpret $A$ and $B$ to be two orbitals lying on a single site.
Transforming to momentum space, $c_{n} = \frac{1}{\sqrt{N}} \sum_k e^{ik \cdot n a} c_{k},$ we have 
\begin{align}
    H_{SSH} &= \sum_k \begin{pmatrix} c_{A,k}^\dagger & c_{B,k}^\dagger \end{pmatrix} 
h_{SSH}(k)
\begin{pmatrix} 
c_{A,k} \\ 
c_{B,k} 
\end{pmatrix},\nonumber\\
h_{SSH}(k)&\equiv\begin{pmatrix} 
0 & t + \lambda e^{-ik} \\ 
t + \lambda e^{ik} & 0 
\end{pmatrix}\\
&=[t + \lambda \cos(k)]\sigma^1 + \lambda \sin(k)\sigma^2 .
\end{align}
As there is no intra-sublattice hopping, 
there is a chiral symmetry that acts as $\left\{h_{ssh}(k),\sigma^3\right\}=0$.
Furthermore, there is an inversion symmetry $\mathcal{I}=\sigma^1$.
Depending on the relative magnitude of $t$ and $\lambda$, the phases have localized Wannier states on different bonds of the dimerized chain, as shown in Fig.~\ref{fig:ssh_fig}.
The critical point occurs at $|t|=|\lambda|$.
To see this by inspection, we can go to the atomic insulating limit, in which one coupling goes to $\infty$.
In that case, we can access two different atomic insulators, one with bond $A$-centered charge and
one with bond $B$-centered charge.
To smoothly deform between these two states would require moving the center of
the localized wavefunction away from an inversion center, breaking the inversion symmetry of the system.

As will also be true for the higher dimensional models, there are many invariants which can be used to sharply distinguish the two phases--these include inversion eigenvalues of the Wannier states at high symmetry $k$ points, the polarization (Zak-Berry phase)
\cite{berry_proc_1984,zak_prl_1989}, and even the presence of zero modes when the model is placed on a finite system with open boundary conditions.
However, our focus is not on the characterization of each phase; we will instead focus on the critical theory separating the phases.

From the dispersion $\epsilon(k)=\pm\sqrt{t^2+\lambda^2+2t\lambda\cos(k)}=0$, we see the bulk band gap closes at the critical point. Let us focus near the critical point $t=\lambda$. Writing $H(k)=( t+\lambda\cos k)\sigma^1+\lambda \sin k \sigma^2$, near the gap closing at $k=\pi$, we obtain an effective Dirac theory in the continuum limit,
\begin{align}
   h_{SSH}(k = \pi + q) &\approx (t - \lambda) \sigma_x - \lambda q \sigma_y\nonumber
   \\
   \implies \mathcal{H}_{crit}
   &=N\int \frac{dq}{2\pi} c_q^{\dagger}h_{SSH}(k=\pi +q)c_q\nonumber
   \\
   \label{eq:ssh_dirac}
   &=\int dx\; \Psi^{\dagger}(m\sigma^1-iv\sigma^2\del_x)\Psi,
\end{align}
where we have defined the continuum field $c_q=\frac{1}{\sqrt{N}}\int dx\; e^{-iqx}\Psi(x)$, mass $m=t-\lambda$, and velocity $v=\lambda$.
Therefore, the critical theory is that of a $(1+1)d$ massless Dirac fermion.

\begin{figure}[b]
\captionsetup{justification=raggedright}
{\includegraphics[width=0.8\columnwidth]{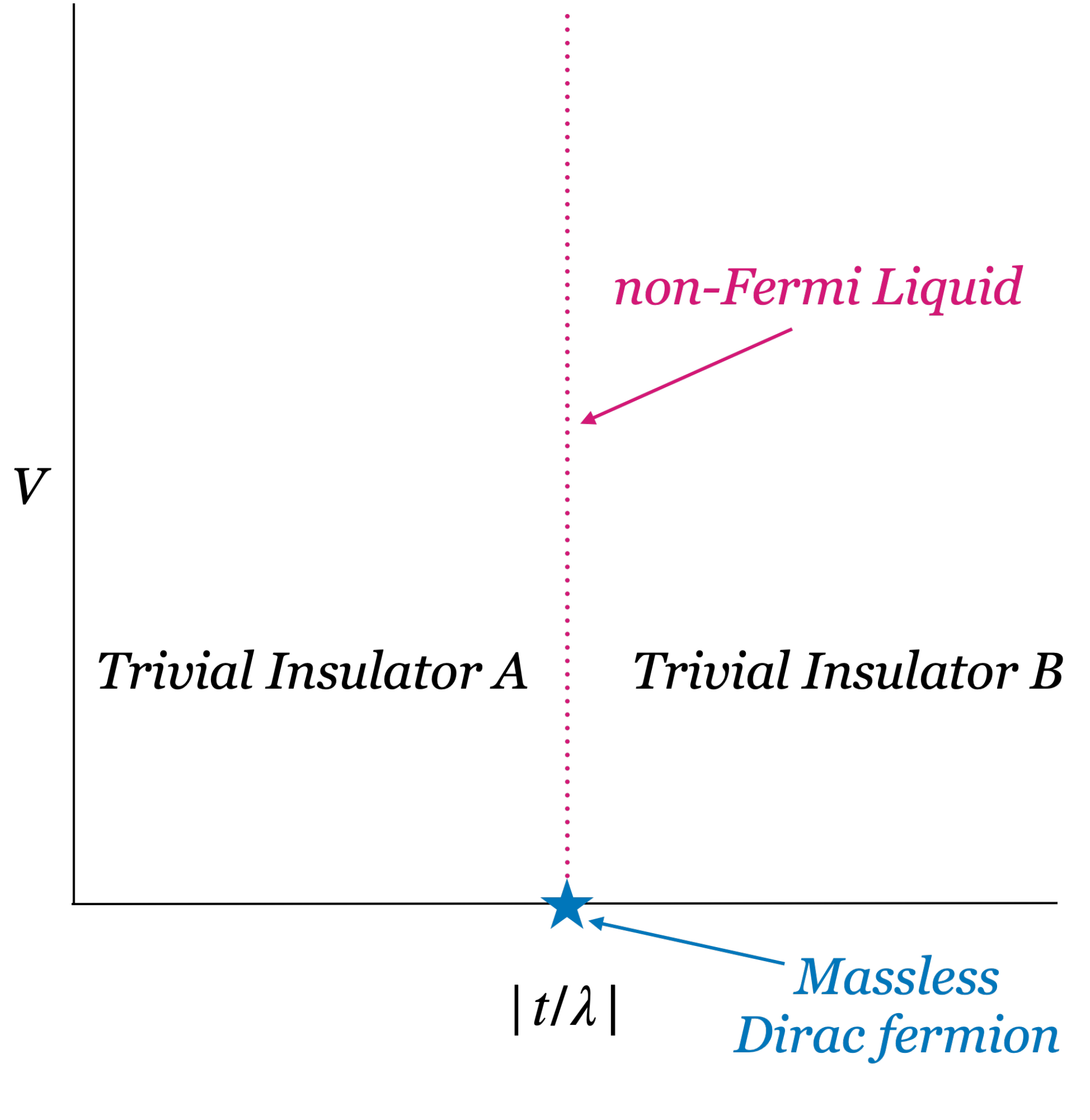}}
\caption{
The phase diagram of the SSH chain the presence of interaction.
Upon adding interactions, there is a marginal line of non-Fermi liquids separating the two trivial insulating phases.
The line terminates at the free fermion critical point in Eq.~\eqref{eq:ssh_dirac}.}
\label{fig:ssh_phase}
\end{figure}

\begin{figure}[h]
\centering
\subcaptionbox{}
{\includegraphics[width=0.9\columnwidth]{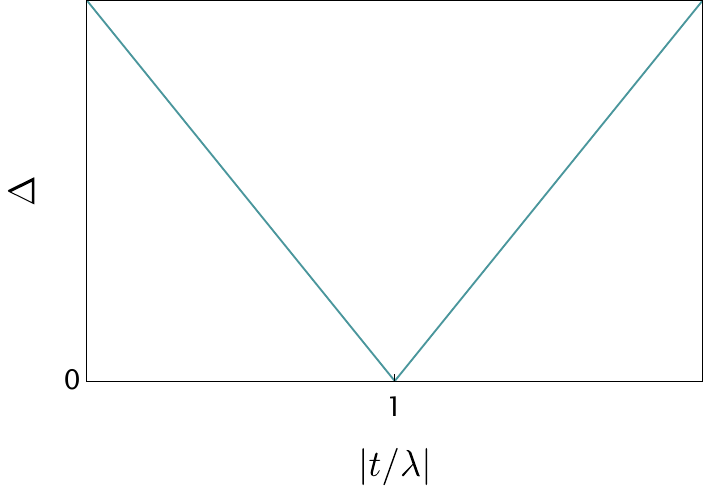}}
\subcaptionbox{}
{\includegraphics[width=\columnwidth]{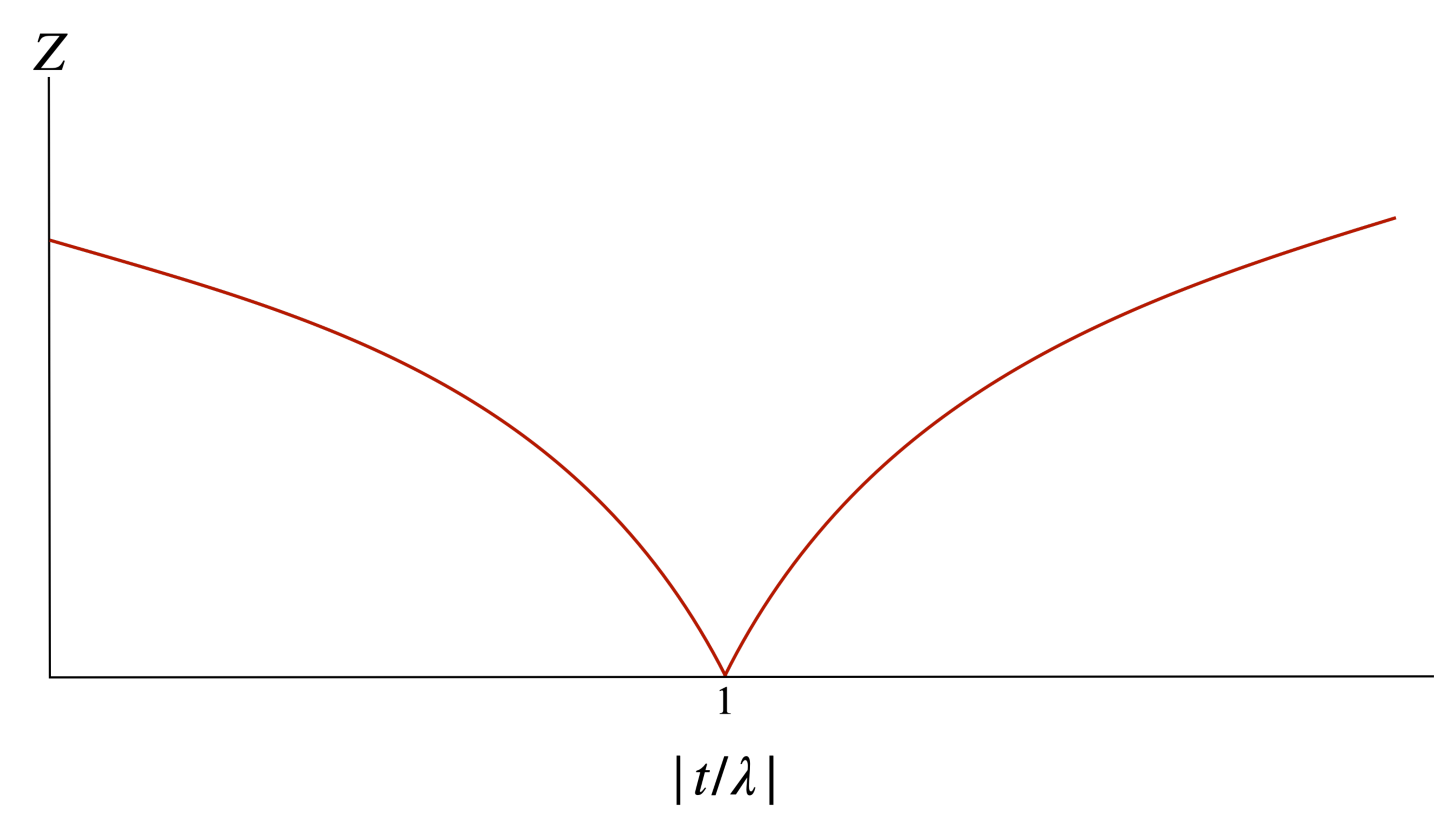}}
\captionsetup{justification=raggedright}
\caption{(a) The vanishing of the energy gap, as the critical point is reached. The gap is is linear in $\delta t\sim t-t_*$.
(b)
Schematic vanishing of the quasiparticle weight as the critical point is approached in the presence of interactions.}\label{fig:ssh_weight}
\end{figure}

Once weak, repulsive interactions are added, the critical theory dramatically changes.
At the quantum critical point, weak interactions are exactly marginal and lead to a renormalization group (RG) fixed line of (non-Fermi) Tomonaga-Luttinger liquids  \cite{haldane_physC_1981,voit_rep_1995}, parametrized by the interaction strength, as shown in Fig.~\ref{fig:ssh_phase}.
Remarkably, the quasiparticle is lost at the quantum critical point even though both neighboring phases are smoothly connected to band insulators.
This is shown in Fig.~\ref{fig:ssh_weight}(a-b), where at the critical point, the vanishing of the energy gap is accompanied by a simultaneous vanishing of the quasiparticle residue $Z\rightarrow 0$.
Furthermore, the spectral function acquires power-law singularities as $\omega$ approaches the Fermi level, as opposed to being described by a $\delta$-function in the non-interacting limit.
In the original fermion Green's function, this is reflected in the presence of branch cuts instead simple poles.
We note the dramatic change of the spectral function from having $\delta$-function behavior to exhibiting 
a power law singularity in the presence of interactions is special to one-dimensional Luttinger liquids; from standard finite density Fermi liquid theory, in higher dimensions, weak interactions
merely broaden the $\delta$-function peak to a Lorentzian and do not destroy the quasiparticle.

\subsection{Bosonic}
We now pivot to the bosonic SSH model, in which the critical point is once again described by a Luttinger liquid.
We can take
\begin{equation}
    H = \sum_{n=1}^{N} \left( t b_{A,n}^\dagger b_{B,n} + \lambda b_{B,n}^\dagger b_{A,n+1} \right) + h.c.,
\end{equation} 
where $b$ is a hard-core boson operator.
We set the filling $\nu=1/2$, or one boson per unit cell.
By previous arguments, there will be a phase transition at $t=\lambda$. 
To elucidate the ground state properties, we perform a Jordan-Wigner transform,
\begin{equation}
        b_i=\exp\left(i\pi\sum_{j=1}^{i-1}c_j^{\dagger}c_j\right)c_i^{} ,
\end{equation}
from which we obtain
\begin{equation}
    H = \sum_{n=1}^{N} \left( t c_{A,n}^\dagger c_{B,n} + \lambda c_{B,n}^\dagger c_{A,n+1} \right) + h.c.,
\end{equation}
which is simply the fermionic SSH.
To characterize the critical point, we can go near the metallic critical point and re-bosonize using standard methods \cite{senechal2004},
\begin{equation}
c_x \approx \sqrt{a}\left( e^{i k_F x} \psi_R(x) + e^{-i k_F x} \psi_L(x) \right),
\end{equation}
where we have defined the chiral fermion fields as vertex operators of bosonic fields $\phi$ and $\theta$
\begin{align}
\psi_R(x) &\sim \frac{1}{\sqrt{2\pi \alpha}} e^{i(\phi(x) + \theta(x))},\\
\psi_L(x) &\sim \frac{1}{\sqrt{2\pi \alpha}} e^{i(-\phi(x) + \theta(x))}.
\end{align}
The bosonic fields satisfy the commutation relations \begin{align}
    [\phi(x), \theta(x')] &= i\frac{\pi}{2} \operatorname{sgn}(x - x'),\\
[\phi(x), \phi(x')] &= [\theta(x), \theta(x')] = 0.
\end{align}
As we have $\psi^\dagger(x) \psi(x) \sim \frac{1}{\pi} \partial_x \phi(x),
$ we can write the Jordan Wigner string as 
\begin{equation}
\exp\left(i\pi \int_{-\infty}^x \psi(y)^{\dagger}\psi(y) \, dy\right) = \exp\left(i k_Fx+i\phi(x)\right).
\end{equation}
Then, using that the original boson can be written in terms of the vertex operators
\begin{equation}
    b(x) \sim \frac{1}{\sqrt{2\pi a}} \left(  e^{i \theta(x)}+e^{2i k_F x} e^{i(2\phi(x) + \theta(x))} +\cdots \right),
\end{equation}
we can write the boson-boson correlator as

\begin{equation}
    \langle b^{\dagger}(x,t)b(0,0)\rangle\sim \frac{1}{2\pi a}\left(F_{0,1}(x,t)+F_{2,1}(x,t)+\cdots\right).
\end{equation}
We have defined 
\begin{widetext}
\begin{equation}
\label{eq:luttinger_greensfunction}
    F_{a,b}(x,t)=\langle e^{i(a\phi(x)+b\theta(x))}e^{-i(a\phi(0)+b\theta(0))}\rangle\sim\frac{\alpha^{(a^2K+b^2/K)/2}}
{(vt-x-i\alpha\Theta(t))^{(a\sqrt{K}-b/\sqrt{K})^2/4}(vt+x-i\alpha\Theta(t))^{(a\sqrt{K}+b/\sqrt{K})^2/4}}
\end{equation}    
\end{widetext}    
in terms of the Luttinger parameter $K$.
At large distances and times, one can see algebraic decay of the correlation function, as expected for a Luttinger liquid.
Therefore, as in the fermionic case, the bosonic SSH critical point is a gapless, non-Fermi liquid (and CFT).

The critical theory obtained by re-bosonizing after a Jordan-Wigner transformation can be reproduced using a parton decomposition. 
We can do this by fractionalizing the boson $b=d_1^{\dagger}d_2$ into fermionic partons $d_{1,2}$.
Performing a mean field decoupling, we can obtain an effective quadratic mean field Hamiltonian,

\begin{equation}
    H_{SSH}(b^{\dagger},b)=H_{SSH}(d_1^{\dagger},d_1)+H_{SSH}(d_2^{\dagger},d_2),
\end{equation}
where each fermionic parton separately realizes an SSH model.
Note 
that the fractionalization of $b$ introduces an emergent $SU(2)$ gauge field that is Higgsed to $U(1)$ by the mean-field decoupling ansatz. 
To project back into the physical Hilbert space, we impose the constraints $d_{1,i}^{\dagger}d_{1,i}+d_{2,i}^{\dagger}d_{2,i}=1$ and $d_{1,i}^{\dagger}d_{1,i}=d_{2,i}^{\dagger}d_{2,i}=\nu_B=\frac{1}{2}$ on all sites $i$.

    The resulting critical theory is $N_f=2$ QED$_{2}$, as there are two Dirac fermions coupled to a $U(1)$ gauge field. 
The low energy limit of this theory is known \cite{gepner_nuclphysb_1985,affleck_nuclphysb_1986,dempsey_prl_2024} and comprises a massive sector in addition to a gapless sector described by the $SU(2)_1$
Wess-Zumino-Witten theory, a CFT with central charge $c=1$. Within abelian bosonization, after bosonizing the fermions $d_{1,2}$ to scalar fields $\phi_{1,2}$, $\phi_1+\phi_2$ acquires a mass, while $\phi_1-\phi_2$ remains massless.
However, there is no emergent $SU(2)$ symmetry; generically, from microscopic perturbations, there is only a $U(1) \times U(1)$ symmetry, of which a diagonal subgroup is the microscopic charge conservation.

\subsection{The Luttinger Liquid Critical Point}
From our discussion, we remark that both the fermionic and bosonic SSH models are described under a unified framework by the sine-Gordon theory for the Luttinger liquid,
\begin{equation}
    \mathcal{L}=\frac{1}{2\pi K v_F}\left(
    (\del_\tau \phi)^2+v_F^2(\nabla \phi)^2\right)-\lambda \cos(2\phi)+\cdots,
\end{equation}
in which the phase transition is tuned by the sign of $\lambda$ and ``$\cdots$'' contains higher harmonic operators.
Note that from the scaling dimensions $\Delta[\cos(2n\phi)]=Kn^2$, we obtain the flow
\begin{equation}
    \dfrac{d\lambda_n}{dl}=(2-Kn^2)\lambda_n
\end{equation} 
under renormalization for couplings $\lambda_{n}\cos(2n\phi)$.
If the Luttinger coefficient $K$ is such that the higher harmonic interactions are irrelevant, then $\lambda=0$ is a quantum critical point.
For the fermionic SSH model,
the electron operator is given by $e^{i(\theta\pm\phi)}$, while for the bosonic SSH model, the boson operator is given by $e^{i\theta}$.
In both cases, we see that in the presence of interactions ($K\ne 1$), there are no quasiparticles at criticality.
From Eq.~\eqref{eq:luttinger_greensfunction}, the simple pole in the free limit, $K=1$, transforms into a branch cut when $K\ne 1$.

\section{Phase Transitions between Atomic Insulators in (2+1)d}
\label{sec:bbh}
\begin{figure}[h]
\captionsetup{justification=raggedright}
{\includegraphics[width=\columnwidth]{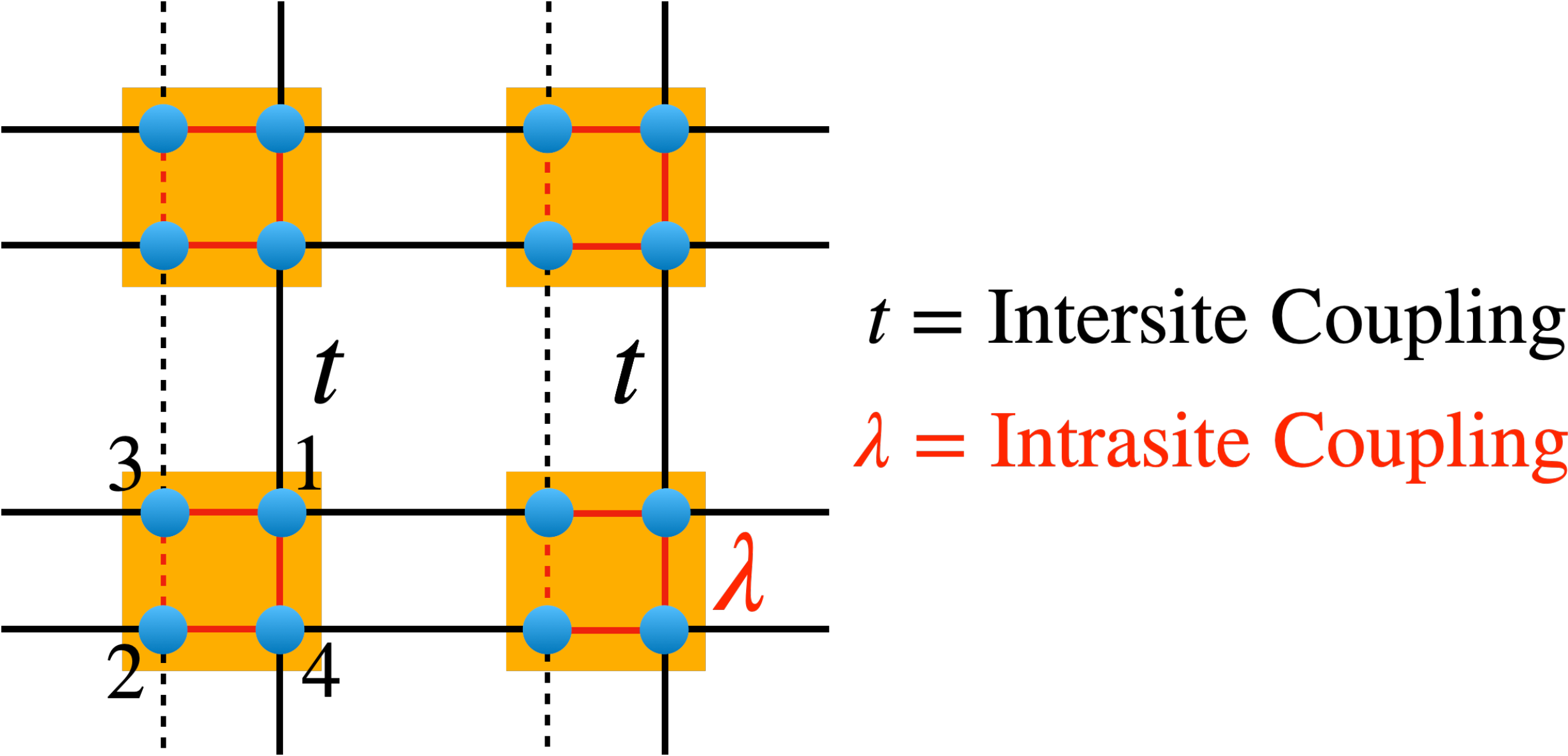}}
\caption{
An illustration of the BBH model. There are four orbitals per site, with a $C_4$ symmetric hopping between orbitals and sites.
The dashed lines above indicate a negative hopping amplitude, a gauge choice that inserts $\pi$ through each plaquette. 
Equivalently, one can view the orbitals as sublattice sites and treat the system as a decorated square lattice.}
\label{fig:bbh_lattice}
\end{figure}

To study an obstructed atomic insulator transition in two dimensions, we take the BBH model \cite{bbh_sci_2017,bbh_prb_2017}, a representative model of a higher order topological insulator, on the square lattice (Fig.~\ref{fig:bbh_lattice}).
The $C_4$ symmetric fermionic model realizes a phase transition between an obstructed atomic insulator with a bulk quadrupole moment and a trivial phase.
The obstructed phase exhibits mirror-symmetry protected midgap corner modes on an open geometry.

As mentioned earlier, while this wider class of multipole insulators can be characterized based on the nested Wilson loop formalism \cite{bbh_sci_2017}, we will focus only on the critical theories of their phase transitions.

In the fermionic model, the phase transition occurs from Dirac bands touching, leading to a bulk gap closing.
The critical theory is that of two massless Dirac fermions, in which short range interactions are irrelevant.

In the bosonic case, fractionalization leads to an emergent QED$_3$ critical point due to the introduction of an emergent $U(1)$ gauge field.
However, the 
proliferation of (renormalization group relevant) symmetry-allowed monopole operators at the critical point leads the critical point to be fine-tuned and ultimately unstable. 
Importantly, the symmetry-allowed monopole operators arise from the bipartiteness of the original square lattice \cite{song_prx_2020}, which enforces a specific embedding of the UV lattice symmetries into the IR that causes the critical point to be fine-tuned.
Consequently, the resulting phase transition is likely first order, and we relegate the details of this model and phase transition to Appendix~\ref{app:bbh}.

\begin{figure}[h]
\captionsetup{justification=raggedright}
\centering
\subcaptionbox{}
{\includegraphics[width=0.7\columnwidth]{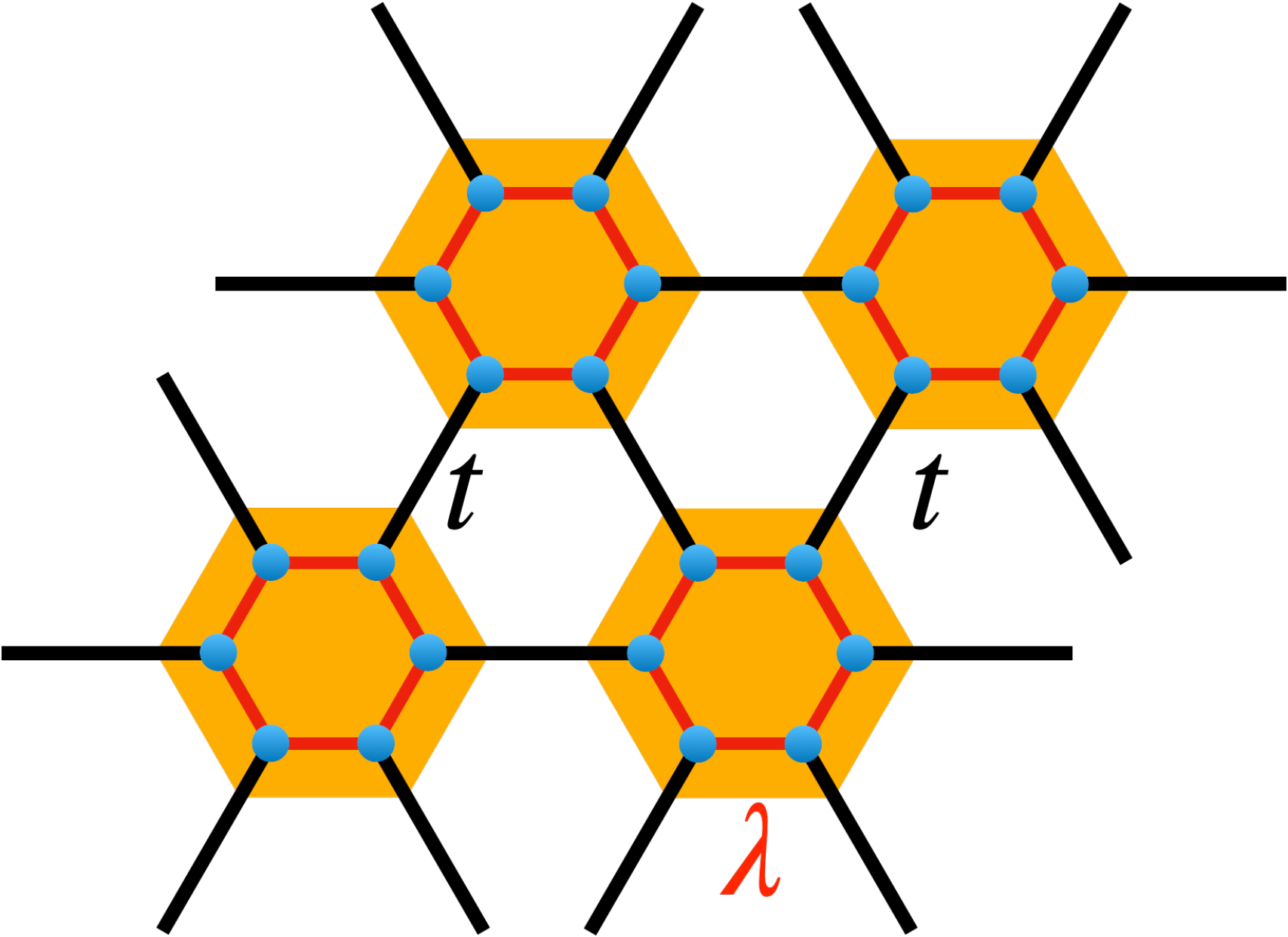}}
\subcaptionbox{}
{\includegraphics[width=0.7\columnwidth]{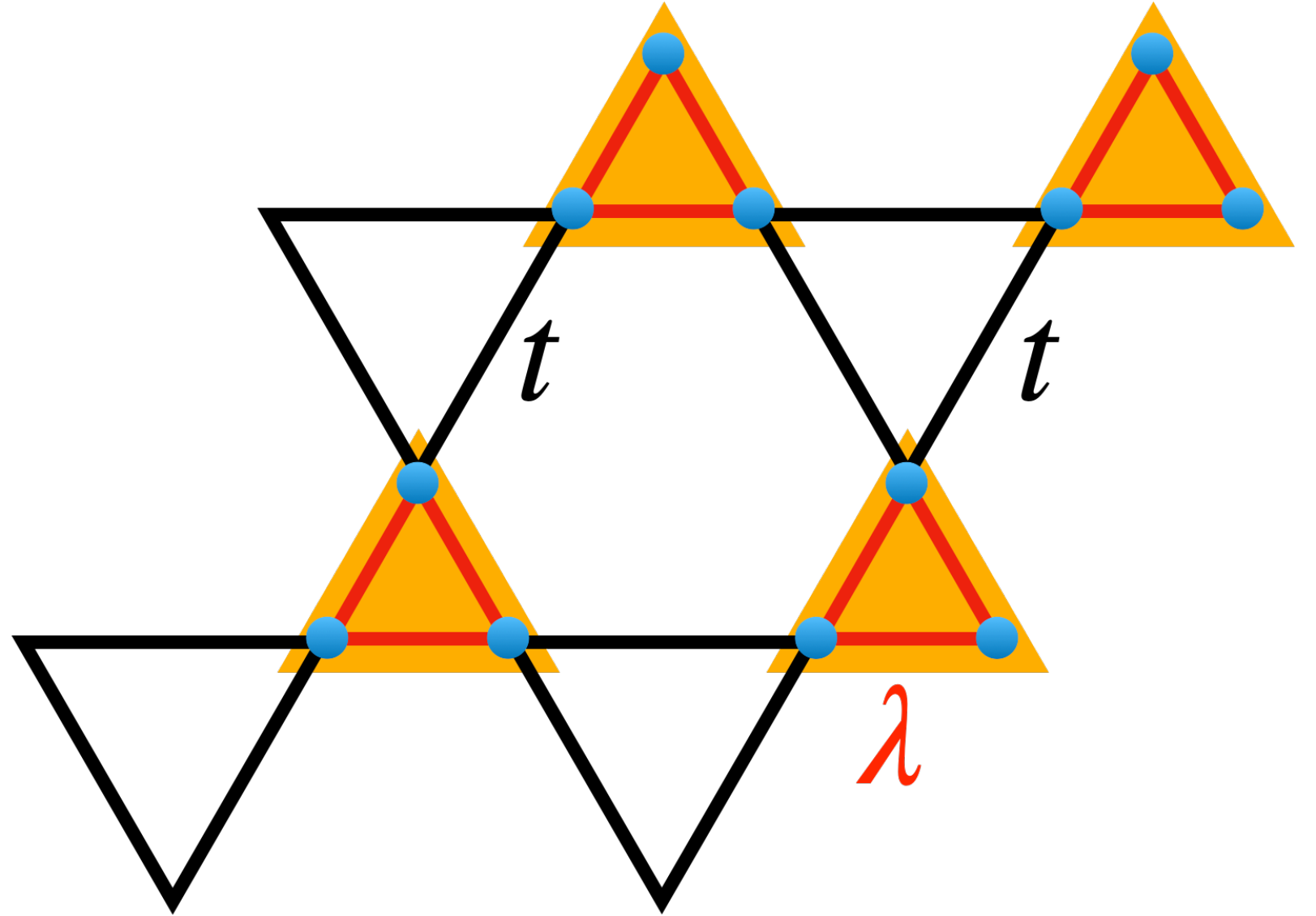}}
\caption{
An illustration of the decorated triangular lattices in the (a) $C_6$ and (b) $C_3$ symmetric cases. As before, $\lambda$ labels the intrasite coupling while $t$ labels the intersite coupling.}
\label{fig:tri_lattices}
\end{figure}

For the analogous atomic insulators on the triangular lattice, we can realize atomic insulators with $C_6$ or $C_3$ rotation symmetry.
The first case is realized by a six-orbital model on the triangular lattice. 
By viewing the orbitals as their own lattice sites, it can be equivalently viewed as a tight binding model on the breathing honeycomb lattice.
The second case is realized by a three-orbital model on the triangular lattice, or equivalently, a tight binding model on the breathing kagome lattice \cite{ezawa_prl_2018}. 
Both lattices are shown in Fig.~\ref{fig:tri_lattices}.

From the bipartite nature of the $C_6$ symmetric model, we expect (and confirm) that the critical point displays similar phenomena to that of the square lattice.
The analysis of this model is detailed in Appendix~\ref{app:c6}.

In the $C_3$ symmetric insulator, however, the lack of bipartiteness allows a possiblity of a stable QED$_3$ critical point and a continuous phase transition. We will explore this model in in the following subsection.
We remark that both the $C_3$ and $C_6$ symmetric models are primitive generators of rotation symmetric higher order topological crystalline insulators in $(2+1)d$ \cite{hughes_prb_2019}.

\subsection{Fermionic Atomic Insulator on the Breathing Kagome Lattice}
\begin{figure*}[t]
\captionsetup{justification=raggedright}
\centering
\subcaptionbox{}
{\includegraphics[width=2\columnwidth]{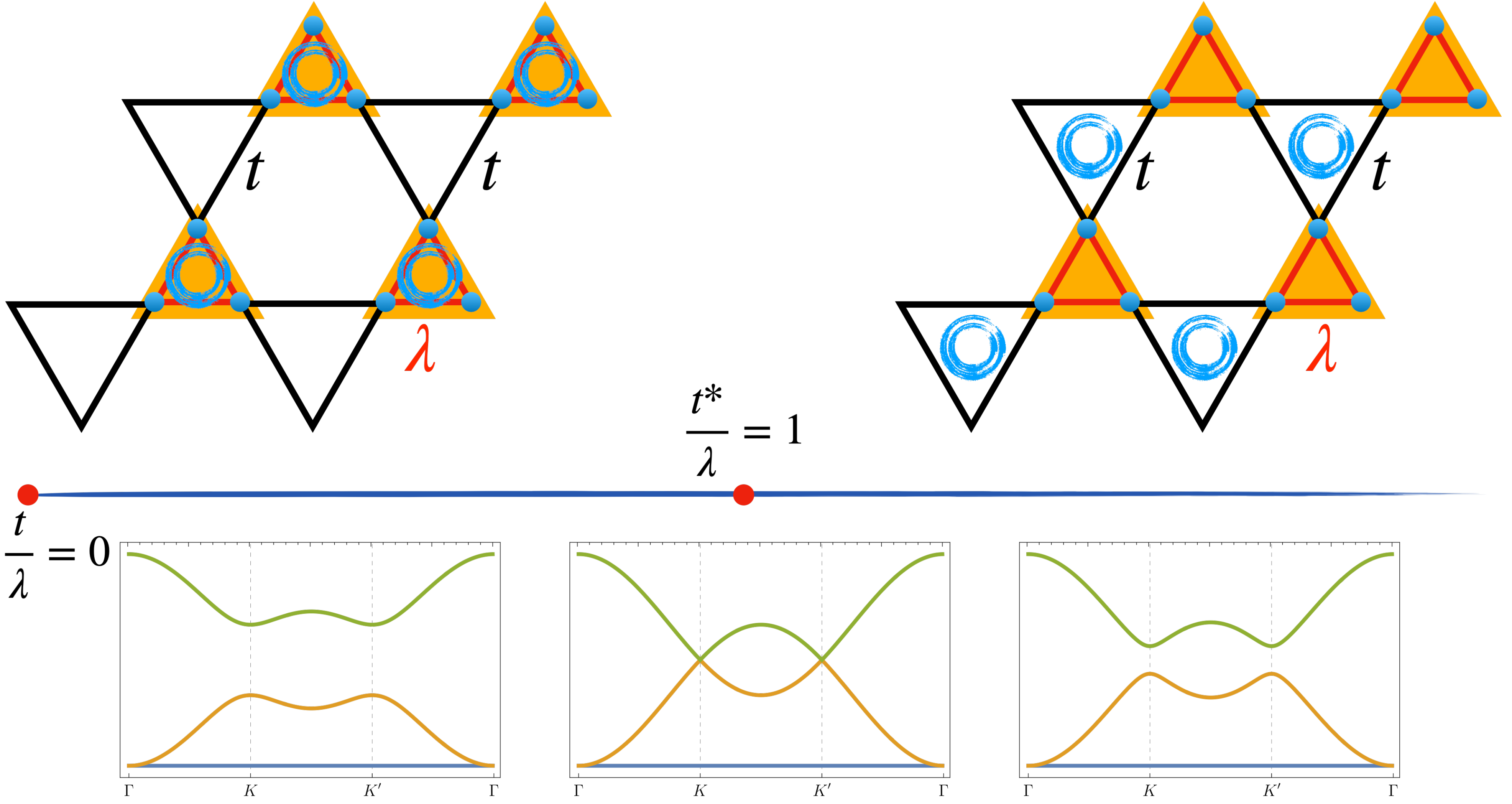}}
\caption{Phase diagram and band dispersions for the $C_3$ symmetric atomic insulator on the breathing kagome lattice. 
The filled in blue circles indicate the orbitals/sites, while the hollow blue circles indicate the particles.
}
\label{fig:kag_phasediag}
\end{figure*}

On the breathing kagome lattice, we can take hopping parameters as in Fig.~\ref{fig:tri_lattices}(b).
For $\lambda>t$, the insulating phase has Wannier orbitals at the center of each triangle, while for $t>\lambda$, the phase has Wannier orbitals centered at each hexagonal plaquette.
At $t=\lambda=1$, the resulting state has Dirac cones at the $C_3$ invariant momenta $\vec{K},\vec{K}'=\pm (\frac{4\pi}{3},0)$.
Unlike in the previous models, this occurs at $2/3$ filling per orbital ($2$ per unit cell).
The relevant symmetries in this case are translation $T_{1,2}$, reflection $\mathcal{R}_x$, time reversal $\mathcal{T}$, and $C_3$ rotation.
Near the critical point, the effective Hamiltonian near the Dirac points is (in units of $t=1$)
\begin{align}
        H_{eff}=&\Psi_{\alpha}^{\dagger}(\vec{q})\left[\tilde{q}^1\tau^1+\tilde{q}^2\tau^2+m\mu^3\tau^3\right]\Psi_{\alpha}(\vec{q})\label{eq:dirac_c3},
\end{align}
where $m=\frac{3}{2}(1-\lambda)$ and $\tilde{q}_{1,2}=\frac{\sqrt{3}}{2}q_{1,2}$.
The symmetries forbid any other mass term, so the Dirac theory describes the critical point.
The phase diagram is shown in Fig.~\ref{fig:kag_phasediag}.

\subsection{Bosonic Atomic Insulator on the Breathing
Kagome Lattice}
As in the one-dimensional case, we will take a single species of hardcore boson $b$ 
hopping on the lattice, and fractionalize it in terms
of fermionic partons $d_{1,2}$,
\begin{equation}
    b=d_1^{\dagger}d_2.
\end{equation}

This rewriting reproduces the physical Hilbert space given we impose the constraint $d_1^{\dagger}d_1+d_2^{\dagger}d_2=1$ at each site/orbital and set the filling $\nu_b=\nu_{d_2}$.
We will only impose the constraints on average to arrive at a mean field Hamiltonian
\begin{equation}
H_{MF}(b^{\dagger},b)=H_{ins,t_1}(d_1^{\dagger},d_1)+H_{ins,t_2}(d_2^{\dagger},d_2),
\end{equation}
where $H_{ins,t_1}$ is the hopping Hamiltonian for the fermionic insulator on the breathing kagome lattice with coupling $t_1$.
Note here we have not specified the form of the original bosonic Hamiltonian.
As long as it respects all the symmetries of the breathing kagome lattice in addition to time reversal, it can take a fully general form.
However, we do assume that there is some parton mean field decoupling 
such that the mean field state above is realized.

As the spectrum is gapless at $2/3$ filling per site (or $\nu_{unit\;cell}=2$), we will fill $\nu_1=1-\nu_2=\frac{1}{3}$, so the mean field $\nu_1+\nu_2=1$ is satisfied.
Mapping the bosons to spin operators, this construction realizes a vacuum with a nonzero spin polarization, explicitly breaking the $SU(2)$ pseudospin symmetry down to $U(1)$.
To ensure the partons simultaneously go critical, we will enforce the time reversal symmetry
\begin{equation}
    \mathcal{T}\;:\;\begin{pmatrix}
        d_1\\d_2^{\dagger}
    \end{pmatrix}\rightarrow \begin{pmatrix}
        -d_2^{\dagger}\\d_1
    \end{pmatrix},
\end{equation}
under which $b$ is invariant.
Then, symmetry under $\mathcal{T}$ requires that $d_1$ and $d_2$ realize the Hamiltonian in Fig.~\ref{fig:tri_lattices}(b) with opposite hopping signs.
The resulting effective theory is then two copies of Eq.~\eqref{eq:dirac_c3} coupled to an emergent $U(1)$ gauge field $a$ and can be described by
$N_f=4$ QED$_3$,
\begin{equation}
\label{eq:qed3_lagrangian}
\mathcal{L}_E=\sum_{N_f=4}\ol{\Psi}(-i\gamma^{\mu}(\del_{\mu}+ia_{\mu}))\Psi+\frac{1}{4e^2}f_{\mu\nu}^2+\delta \mathcal{L}_E.
\end{equation} 
We have defined
$\gamma^{\mu}=(\tau^3,-\tau^2,\tau^1)$ and $\ol{\Psi}=i\Psi^{\dagger}\tau^3$.
The terms in $\delta \mathcal{L}_E$ are needed in order to break the $SU(2)_g$ gauge symmetry of the partons down to $U(1)_g$.
Concretely, $\delta \mathcal{L}_E$ will contain velocity anisotropies coming from imaginary second-nearest neighbor hoppings (this point is discussed in more detail in Appendix~\ref{app:bbh}) and other symmetry-allowed operators.

We remark that the mass in Eq.~\eqref{eq:dirac_c3}, which has been tuned to zero at the critical point, corresponds to a quantum spin and valley Hall mass in the parton picture, $\Psi^{\dagger}m\tau^3\mu^3\sigma^3\Psi$.
In the IR QED$_3$ theory, the UV lattice symmetries act as
\begin{align}
    T_{1}\;:\;&\Psi\rightarrow e^{i\frac{4\pi }{3}\mu^3}\Psi,\\\mathcal{R}_{x}\;:\;&\Psi\rightarrow \frac{i\tau^2}{2}(\mu^2+\sqrt{3}\mu^1)\sigma^3\Psi,\\
    C_3\;:\;&\Psi\rightarrow e^{i\frac{\pi }{3}\tau^3}e^{-i\frac{\pi }{3}\mu^3}\Psi,\\
    \mathcal{T}\;:\;&\Psi\rightarrow \tau^3\sigma^2\Psi^*,
\end{align}
where we have defined translation $T_1$ ($T_2$ acts trivially on $\Psi$), reflection $\mathcal{R}_x$ taking $x\rightarrow -x$, and $C_3$ rotation in addition to time reversal $\mathcal{T}$.

\subsubsection{A Review of the Critical Theory}
The quantum field theory in Eq.~\eqref{eq:qed3_lagrangian} has been studied extensively, and we will briefly review the relevant details here.
We first remark that without $\delta{\mathcal{L}}_E$, it is believed that Eq.~\eqref{eq:qed3_lagrangian}
flows to an interacting CFT fixed point \cite{Appelquist_prl_1988,chesterjhep2016,li_2022_jhep,poland_prd_2022,he_scipost_2022,li2022plb} for $N_f\gtrsim 4$.
Therefore, to analyze the QED$_3$ critical point, one must classify the symmetry-allowed relevant operators contained in $\delta\mathcal{L}_E$.

As the Dirac fermion $\psi$ is not a local observable, the most important class of local operators are monopole operators $\mathcal{M}_q$, which insert an integer $q$ units of $U(1)$ gauge flux at a spacetime point and describe topologically nontrivial configurations of the gauge field \cite{borokhov_jhep_2022,borokhov_jhep_2022b}. 
The monopole operators are charged under the $U(1)_{top}$ conserved current
\begin{equation}
        j^{\mu}=\frac{1}{2\pi}\epsilon^{\mu\nu\lambda}\del_{\nu}a_{\lambda}.
\end{equation}
In the presence of $N_f$ Dirac fermions, the monopole operators $\mathcal{M}_q$ must be dressed by fermion zero modes in order to be gauge invariant. 
The fundamental monopole operators, $\mathcal{M}_{q=\pm 1}$, carry unit flux and must have half of its available zero modes filled. As each Dirac fermion contributes a single zero mode in a $\pm 2\pi$ flux background, we see there is a total of $C(4,2)=6$ fundamental monopoles. 

Schematically, we can write the fundamental monopoles as
\begin{equation}
    \phi^{\dagger}\sim f_{i}^{\dagger}f_{j}^{\dagger}\mathcal{M}_{1},
\end{equation}
where $f_i^{\dagger}$ is the fermion zero mode associated with $\Psi_i$ and $\mathcal{M}_{1}$ creates a single flux quanta in the spacetime without nucleating any zero mode.

All other local operators, such as fermion billinears, can be constructed as composites of the monopole operators (for a more detailed discussion, one can refer to the Appendix of \cite{zhang_scipost_2025}). 
Therefore,
the monopole operators can be viewed as the core building blocks of the theory.
Furthermore, from a $1/N_f$ expansion, the scaling dimension of the
fundamental monopoles $\phi$ is given by \cite{pufu_prd_2014,dyer_monopole_2013}
\begin{equation}
\Delta_{\phi}= 0.265N_f-0.0383 +\mathcal{O}(1/N_f)\sim 1.02<3,
\end{equation}
which is relevant at the QED$_3$ fixed point.
Barring fermion billinears, we will assume all other operators (including four fermion terms and higher charge monopoles) in the theory to be irrelevant.
These assumptions are within those commonly made for studies of QED$_3$ theories arising from the lattice \cite{he_scipost_2022,zhang_scipost_2025,zhang2025unnecessaryquantumcriticalitysu3}.
Therefore, the fundamental monopoles are the most important operators in determining the ultimate fate of the quantum critical point.
Using the methods developed in \cite{song_prx_2020,song_nat_2019}, we will analyze the quantum numbers of monopole operators.

Before proceeding, we note that Eq.~\eqref{eq:qed3_lagrangian} will also generically contain velocity anisotropy terms.
Refs. \cite{vafek_prb_2002,vafek_prl_2002,hermele_prb_2005,hermele_prb_2005_erratum} have shown a class of these, specified by a valley anisotropy, are irrelevant.
Other velocity anisotropies include intravalley contributions.
Using a large $N_f$ expansion, 
we explicitly show that all velocity anisotropies are irrelevant at the QED$_3$ fixed point in Appendix~\ref{app:aniso}.

\subsubsection{Monopole Quantum Numbers}
We first observe that the QED$_3$ critical point has an enhanced symmetry group compared to that of the atomic insulators on either side of the transition. 
The QED$_3$ critical point has infrared (IR) time reversal, reflection, charge conjugation, and Lorentz spacetime symmetries, in addition to an IR symmetry,
\begin{equation}
    \dfrac{SO(6)\times U(1)_{top}}{\mathbb{Z}_2},
\end{equation}
describing the transformation of the monopole operators, the local operators of the theory.
The $SO(6)=SU(4)/\mathbb{Z}_2$ symmetry above arises from the $SU(4)$ fermion flavor symmetry taking $\psi_i\rightarrow U_{ij}\psi_j$.
The common $\mathbb{Z}_2$ quotient arises from the fact that, for all local operators, a $\pi$ rotation in $U(1)_{top}$ can be compensated by a transformation by the $SO(6)$ center $-\mathbb{I}_{6\times6}$.
Specifically, the $6$ fundamental monopoles  transform as a six-dimensional vector under $SO(6)$.

As the $N_f=4$ flavors of $\Psi_i$ naturally split into $2$ valleys along with $2$ fermionic parton pseudospins, for convenenience, we will 
define the six monopoles as
\begin{align}
    \phi^{\dagger}_{1,2,3}&=f^{\dagger}\left(i\mu^2\mu^{1,2,3}\otimes i\sigma^2\right)f^{\dagger}\mathcal{M}_{bare},\\
    \phi^{\dagger}_{4,5,6}&=if^{\dagger}\left(i\mu^{2}\otimes i\sigma^2\sigma^{1,2,3}\right)f^{\dagger}\mathcal{M}_{bare},
\end{align}
where $\sigma$ acts on the pseudospin index of the fermionic partons and as defined earlier, $\mu$ acts on the valley index.
We have set up the monopole basis to respect branching $SO(6)\rightarrow SO(3)_v\times SO(3)_f$ for convenience, as
the UV symmetries will embed in a way that factors through this product.
Note that $\phi_{1,2,3}$ is a singlet under the pseudospin $SO(3)_f$ while $\phi_{4,5,6}$ is a singlet under the valley $SO(3)_v$.
On the Dirac fermions, we will choose the IR symmetries to act as
\begin{align}
\label{eq:ir_discrete_sym}
\mathcal{T}_{IR}\;:\;&\Psi\rightarrow i\gamma^1\mu^2\sigma^2\Psi,\quad i\rightarrow -i,\\
    \mathcal{R}_{x,IR}\;:\;&\Psi\rightarrow i\gamma^1\Psi,\\
    \mathcal{C}_{IR}\;:\;&\Psi\rightarrow i\gamma^1\mu^2\sigma^2
    \Psi^*,
\end{align}
where the subscript $IR$ indicates that these are the IR symmetries.
We note that the UV symmetries coming from the lattice projective symmetry group, when descending to the IR, will usually involve some combination of the continuum IR symmetries composed with suitable $SO(6)\times U(1)_{top}/\mathbb{Z}_2$ transformations, as $f$ transforms like $\psi$ under the flavor symmetry \footnote{Note that there some ambiguity arising from the common $\mathbb{Z}_2$ quotiented from $SO(6)\times U(1)_{top}$ global symmetry.
The $\mathbb{Z}_2$ factor acts as $-1$ on all local operators and can be chosen to be part of $SO(6)$ or $U(1)_{top}$, depending on convention.
This is the same convention that determines whether a symmetry acts as $\psi\rightarrow U\psi$ for $U\in SU(4)$ or $\psi\rightarrow iU$, which differs by a factor of $-1$ on the monopole operators. As the physical symmetry is not convention dependent, the Berry phase should change by $\pi$ between difference choices of $U$ or $iU$.
We fix this element of $SU(4)$ in our calculations by the transformation of the six fermion billinear masses $\{\ol{\psi}\mu^i\psi,\ol{\psi}\sigma^i\psi\}$, which also form a vector representation of $SO(6)$.}.
For example, $\mathcal{T}_{IR}$ reverses the monopole flux and from its action on $\Psi$, one can show it must act as 
\begin{equation}
    \phi\rightarrow \pm M_{\mathcal{T}_{IR}}\phi^{\dagger},\quad M_{\mathcal{T}_{IR}}=\begin{pmatrix}
    \mathbb{I}_{3\times 3}&\\&-\mathbb{I}_{3\times 3}
\end{pmatrix}
\end{equation} on the six monopoles.
The $\pm$ factor arises because though the action of any symmetry on $\Psi$ specifies the $SO(6)$ transformation factor, the monopole Berry phase coming from the $U(1)_{top}$ must be extracted with other methods.
Inherently, the Berry phase is a UV contribution, and intuitively, the phase is accumulated by the monopole moving around
the lattice in the parton charge background.

One can also obtain the action of the IR charge conjugation symmetries and reflection.
From the $SO(6)$ action of charge conjugation, we see it must act on the monopoles in the same way as the IR time reversal up to an undetermined phase. 
As reflection reverses monopole flux, we have that $\mathcal{R}_{x,IR}$ acts as 
$\phi\rightarrow \phi^{\dagger}$,
up to a uniform phase factor $e^{i\theta_{\mathcal{R}_{x,IR}}}$.
We will fix the overall phase of $\phi$ such that $\theta_{\mathcal{R}_{x,IR}}=0$.

To find the Berry phase associated with $\mathcal{T}_{IR}$ and $\mathcal{C}_{IR}\mathcal{R}_{x,IR}$, we follow the analytic argument from \cite{song_prx_2020}.
For simplicity, let us assume the UV lattice system has a full $SO(3)_f$ pseudospin symmetry.
First, we introduce a Dirac quantum spin Hall mass $\ol{\psi}\sigma^3\psi$, which splits the zero mode degeneracy and leaves only the monopole $\Phi\sim\phi_4+i\phi_5$ that transforms as $S_z$ to be gapless.
As the spin Hall mass preserves the IR time reversal, the time reversal Berry phase for the original system can be calculated in the presence of this spin Hall mass.
Then, the nontriviality of the spin Hall state $\mathbb{Z}_2$ topological insulator \cite{kane_prl_2005} is manifested through the monopole acquiring a minus sign under the Kramers time reversal, $\Phi\rightarrow - \Phi^{\dagger}$.
This can be seen by examining the edge theory of the spin Hall insulating state and identifying the monopole with a particular tunneling operator on the edge.
The same arguments apply to $\mathcal{C}_{IR}\mathcal{R}_{x,IR}$, leading to
\begin{align}
\mathcal{T}_{IR}\;:\;
&\begin{pmatrix}
        \phi_{1,2,3}\\
        \phi_{4,5,6}
    \end{pmatrix}
    \rightarrow
    \begin{pmatrix}
        \phi_{1,2,3}^{\dagger}\\
        -\phi_{4,5,6}^{\dagger}
    \end{pmatrix},\nonumber\\
\label{eq:baresym_monopole} 
\mathcal{C}_{IR}\mathcal{R}_{x,IR}\;:\;&\begin{pmatrix}
        \phi_{1,2,3}\\
        \phi_{4,5,6}
    \end{pmatrix}
    \rightarrow
    \begin{pmatrix}
        \phi_{1,2,3}\\
        -\phi_{4,5,6}
    \end{pmatrix}.
\end{align}
Note that while the above argument does not discriminate between the valley and spin $SU(2)$ symmetries, the same argument does not apply when we choose to a condense a valley Hall mass $\ol{\psi}\mu^3\psi$. 
This is because the resulting valley Hall state does not admit a physical edge to the vacuum due to a parity anomaly, manifested as a mixed anomaly between $SO(3)_{f}$ and $SO(3)_{v}$.
Such an anomaly does not affect the spin Hall state because the continuous valley symmetry is not present in the microscopic system and instead is only emergent in the IR. 
Note that, as we will discuss shortly, the above arguments can be generalized to the case of the breathing kagome atomic insulator, which only has a microscopic $U(1)\subset SO(3)_f$ psuedospin symmetry.

We are now ready to  analyze the monopole quantum numbers for the $C_3$ symmetry atomic insulator.
To begin, we will list the final transformation properties of the monopoles in Table~\ref{table:monopole_fund_c3}.
\begin{table*}[t]
\captionsetup{justification=raggedright}
\begin{center}
\begin{tabular}{|c|c|c|c|c|}
\hline
Monopole                 & $T_{1}$              & $C_3$              & $\mathcal{R}_x$             & $\mathcal{T}$     \\ \hline
$\phi_1^{\dagger}$       & $\dfrac{-\phi_1^{\dagger}+\sqrt{3}\phi_2^{\dagger}}{2}$        & $\dfrac{-\phi_1^{\dagger}-\sqrt{3}\phi_2^{\dagger}}{2}$       & $\dfrac{\phi_1^{}-\sqrt{3}\phi_2}{2}$      & $\phi_1^{\dagger}$       \\ 
$\phi_2^{\dagger}$       & $\dfrac{-\sqrt{3}\phi_1^{\dagger}-\phi_2^{\dagger}}{2}$       & $\dfrac{\sqrt{3}\phi_1^{\dagger}-\phi_2^{\dagger}}{2}$       & $\dfrac{-\sqrt{3}\phi_1^{}-\phi_2}{2}$      & $\phi_2^{\dagger}$       \\ 
$\phi_3^{\dagger}$       & $\phi_3^{\dagger}$       & $\phi_3^{\dagger}$        & $-\phi_3^{}$       & $\phi_3^{\dagger}$           \\ 
$\phi_{4/5/6}^{\dagger}$ & $\phi_{4/5/6}^{\dagger}$ & $\phi_{4/5/6}^{\dagger}$ & $-\phi_{4/5}^{},\phi_{6}^{}$ & $-\phi_{4}^{\dagger},\phi_{5}^{\dagger},-\phi_{6}^{\dagger}$ \\ \hline
\end{tabular}
\end{center}
\caption{The transformation of the single-charge monopoles under the UV symmetries in the breathing kagome lattice critical point. Note the monopoles are all trivial under $T_2$ translation.
The above symmetries, combined with $U(1)_b$ number conservation symmetry of the hardcore bosons, which rotates $d_{1,2}\rightarrow e^{i\sigma_3}d_{1,2}$, forbid a trivial monopole.}
\label{table:monopole_fund_c3}
\end{table*}
To arrive at the transformations, we first note that the translations $T_{1,2}$ should act with trivial $U(1)_{top}$ Berry phase. This can be shown algebraically using that
the critical point has an emergent $C_6$ about each hexagonal plaquette.
From the relation $T_1C_6T_2=T_2C_6$ and the fact that $\phi_{4,5,6}$ transform with only a phase under $T_{1,2}$ and $C_6$, one can see that the Berry phase for $T_{1}$ must vanish. 
The Berry phase for $T_2$ must also vanish by the relation $C_6T_1=T_2C_6$.
To find the Berry phase for $C_3$, we can use numerical methods \cite{song_nat_2019}, putting the system on a torus and threading $2\pi$ flux uniformly.
Filling the appropriate zero modes (which will have a finite size gap), we calculate the quantum numbers of the resulting many body state to confirm the valley singlet monopoles have zero $C_3$ angular momentum, as well as zero lattice momentum.

Lastly, we must determine the Berry phases for the discrete symmetries. As before, we set $\mathcal{R}_x$ to have trivial Berry phase as this corresponds to fixing an overall phase shift of all monopoles.
To determine the Berry phase for $\mathcal{T}$, it will be useful to recall the IR QED$_3$ symmetries defined in Eq.~\eqref{eq:ir_discrete_sym}.
We see that $\mathcal{T}$ acts exactly as the continuum $\mathcal{C}_{IR}\mathcal{T}_{IR}$ composed with a spin flip symmetry $i\sigma^2$ and a Lorentz rotation that does not affect the monopoles.
As $\mathcal{R}_{IR}$ acts as $\phi^{\dagger}\rightarrow \phi$, determining the Berry phase associated with $\mathcal{T}_{IR}$ and $\mathcal{C}_{IR}\mathcal{R}_{IR}$ will determine the Berry phase of $\mathcal{T}$.
To do this, we will condense a quantum spin Hall mass $\ol{\psi}\sigma_3\psi$ as before, leading to a nontrivial topological insulator. 
Even though we do not have microscopic $SU(2)$ pseudospin symmetry, there is still a residual $U(1)$ rotation about the pseudospin polarization axis corresponding to the $U(1)_b$ boson conservation symmetry, which is unbroken by the spin hall mass.
Consequently, the resulting state is still a nontrivial $\mathbb{Z}_2$ topological insulator. The previous arguments then still apply, and the $\mathbb{Z}_2$ nontriviality is reflected in the phase of the spin triplet monopole acquiring a minus sign under time reversal.
The final
transformations of the IR discrete symmetries are identical
to the cases in which full $SU(2)$ symmetry is present,
\begin{align}\mathcal{T}_{IR}\;:\;&\begin{pmatrix}
        \phi_{1,2,3}\\
        \phi_{4,5,6}
    \end{pmatrix}
    \rightarrow
    \begin{pmatrix}
        \phi_{1,2,3}^{\dagger}\\
        -\phi_{4,5,6}^{\dagger}
    \end{pmatrix},\\ \mathcal{C}_{IR}\mathcal{R}_{IR}\;:\;&\begin{pmatrix}
        \phi_{1,2,3}\\
        \phi_{4,5,6}
    \end{pmatrix}
    \rightarrow
    \begin{pmatrix}
        \phi_{1,2,3}\\
        -\phi_{4,5,6}
    \end{pmatrix}.
\end{align}
From the above, the monopoles under $\mathcal{T}$ must have the transformation as outlined in Table.~\ref{table:monopole_fund_c3}.
\subsubsection{The bosonic critical point}
The results of Table.~\ref{table:monopole_fund_c3}, along with $U(1)_b$ boson conservation symmetry, forbid the proliferation of any trivial monopole operator, as the only singlet monpoles under the lattice and discrete symmetries are either non-Hermitian (such as $\phi_3^{\dagger}-\phi_3$) or break $U(1)_b$ (such as $\Im[\phi_4]$).

The next most relevant operators are the fermion billinear operators, which we have tuned to zero to reach the critical point.
As discussed earlier, other singlet operators, which include excited charge $1$ monopole operators, higher charge monopoles, and operators which transform as symmetric two-index tensors of $SO(6)$, are likely irrelevant by analytic and numeric computations, or they must be assumed irrelevant if $N_f=4$
QED$_3$ is found to be stable in any lattice gauge model (more detailed discussion in \cite{zhang_scipost_2025}).
Therefore, we 
find that the critical point between the two trivial insulating phases on the breathing kagome lattice is
a natural candidate for realizing the QED$_3$ CFT.
This CFT is stable and unfine-tuned as there is only a single relevant perturbation (the fermion mass) tuning the transition.

\begin{figure}[t]
\captionsetup{justification=raggedright}
{\includegraphics[width=\columnwidth]{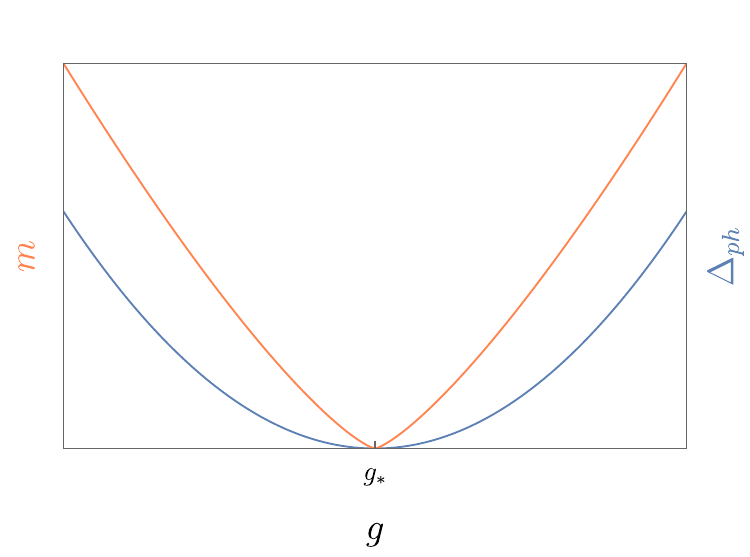}}
\caption{Illustration of the fermion gap $m$, induced by the bilinear mass, and the photon gap $\Delta_{ph}$, induced by monopole proliferation, as the system is tuned across criticality with a general
coupling $g$. Note that we presume that $\Delta_{ph}\sim m^y$ with $y>1$, as it is induced by an operator that is dangerously irrelevant at the QED$_3$ critical point.}
\label{fig:kagome_criticality_gaps}
\end{figure}

The insulators on each side of the transition can be viewed as QED$_3$ perturbed by an $SU(4)$ adjoint mass, which breaks the IR global symmetry to $(SO(4)\times SO(2))/\mathbb{Z}_2$. 
However, note that the appearance of a trivial single-charge monopole is forbidden by the UV symmetries. 
While it is possible that the adjoint mass can lead to monopole proliferation which precipitates spontaneous symmetry breaking \cite{song_nat_2019}, we assume instead that the adjoint mass drives proliferation of higher charge, UV singlet monopoles, which further breaks the IR global symmetry to the UV symmetry of the lattice insulator.
These higher charge monopoles, which drive confinement, are in the class of UV singlet operators that are irrelevant at the critical point, but become relevant once we deform away from criticality.
The schematic scaling of the fermion (parton) gap, $m$, and the photon gap $\Delta_{ph}$ away from the critical point is shown in Fig.~\ref{fig:kagome_criticality_gaps}.
In particular, we note that as the photon is gapped by higher charge, dangerously irrelevant monopoles,
$\Delta_{ph}$ will be parametrically smaller than the fermion gap $m$.

\section{Anomaly Matching Conditions}
\label{sec:lsm_matching}
In this section, we confirm the quantum numbers of the monopoles in the three models we have analyzed is indeed consistent with what is expected from the lattice Lieb-Shultz-Mattis-Oshikawa-Hastings (LSMOH) constraints \cite{lieb_ann_1961,oshikawa_prl_2000,hastings_2004_prb,cheng_prx_2016,po_prl_2017,huang_prb_2017,cho_prb_2017,jian_prb_2018,thorngren_prx_2018}.
Applying the tools developed in \cite{zou_prx_2021,ye_scipost_2022}, this illustrates that the critical theories discussed and the symmetry transformations of the monopoles, including for the unstable multicritical points on the bipartite lattices, are consistent with what can be emergent from a local bosonic lattice Hamiltonian like the ones we described, providing a nontrivial check on our results. 

We will briefly describe the general framework for LSMOH anomaly matching.
To begin, we view the critical QED$_3$ field theory as emergent from the lattice at the gapless critical point.
The local operators in the QED$_3$ theory are then coarse grained version of lattice operators, matched by their symmetry properties.
Formally, the correspondence between the UV and IR symmetries is determined by a group homomorphism from the UV symmetry group ${G}_{UV}$ to the IR symmetry group $G_{IR}=\frac{SO(6)\times U(1)_{top}}{\mathbb{Z}_2}$ (in addition to discrete symmetries), 
\begin{equation}
    \varphi\;:\;G_{UV}\rightarrow G_{IR}.
\end{equation}
However, it turns out that not every symmetry embedding $\varphi$ is physically consistent. 
In particular, if the IR theory contains a 't Hooft anomaly $\eta[G_{IR}]$, that anomaly must pull back under $\varphi$ to the anomaly of the UV,
\begin{equation}
\label{eq:anom_match}
 \eta[G_{UV}]=\varphi^*\eta[G_{IR}].
\end{equation}
As $\eta[G_{UV}]$ is determined by lattice LSMOH theorems, anomaly matching provides a constraint on how $\varphi$ can act. In
Eq.~\eqref{eq:anom_match}, both $\eta[G_{UV}]$ and $\varphi^*\eta[G_{IR}]$ are elements in $H^4(G_{UV},U(1)_T)$, where $U(1)_T$ indicates time reversal acts as complex conjugation.
While Eq.~\eqref{eq:anom_match} is a necessary condition for a theory to be ``emergeable'' from the lattice, it is expected that anomaly matching is also a sufficient condition.

The most simple LSMOH theorem applies to a lattice with an odd number of spin-$1/2$ degree of freedoms in each translation invariant unit cell.
Upon coupling the
system to background gauge fields of the $SO(3)$ spin symmetry and the $T_{1,2}$ translation symmetries, the resulting theory is anomalous, with a bulk SPT anomaly 
\begin{equation}
    \eta[G_{UV}]=\exp\left(i\pi \int_{X_4}w_2^{SO(3)}\cup x \cup y\right),
\end{equation}
where $x,y\in H^1(X_4,\mathbb{Z}_2)$ are the translation gauge fields.
There are similar anomalies that activate when an odd number of spin-$1/2$ moments located on reflection or rotation axes.
Furthermore, there are LSMOH constraints for time reversal when the spin-$1/2$ object is also a Kramers doublet.

All the models we consider are particularly simple, in that there is no lattice LSMOH anomaly. Therefore, all of the IR anomalies should pullback to zero and the LHS of Eq.~\eqref{eq:anom_match} is trivial. 
However, as the RHS of Eq.~\eqref{eq:anom_match} describes the nontrivial IR anomaly, the matching condition still enforces nontrivial relations for the symmetry embedding $\varphi$.
To explicitly show this, we first write down the IR anomaly for $N_f=4$ QED$_3$ by coupling $SO(6)$ and $U(1)_{top}=SO(2)$ gauge fields, considering it as a Stiefel liquid \cite{zou_prx_2021}.
The resulting nonlinear sigma model has the same symmetries and anomalies as QED$_3$.

We will perform the proper anomaly matching procedure in Appendix~\ref{app:lsm}. In particular, simplification occurs because of
the absence of UV lattice LSM anomalies in the critical points we consider.

Note that anomaly matching does not make any statement about the dynamics of the IR theory.
In particular, for the IR theory to be stable, all relevant perturbations must not be singlets under $G_{UV}$.
For the case of bipartite lattices, while anomaly matching shows QED$_3$ is an emergeable CFT at the critical point, our earlier analysis shows that it is dynamically unstable due to the presence of RG relevant $G_{UV}$ singlet monopoles.

\section{Outlook}
\label{sec:outlook}
In this paper, we have examined how a certain class of phase transitions between trivial bosonic atomic insulating states can realize exotic critical points through fractionalization.
Beginning with the SSH chain as a warm-up example, we showed how both fermionic and bosonic systems lead naturally to a Luttinger liquid at criticality, in which the underlying particles do not have integrity as quasiparticles. 
Elevating to two dimensions via the BBH model on the square lattice, a bosonic parton construction that fractionalizes the boson into two fermions produces an $N_f=4$ QED$_3$ description of the putative critical point. 
Crucially, however, lattice symmetries control the monopole content of the low-energy theory--on bipartite lattices, a symmetry-allowed monopole operator exists and destabilizes the QED$_3$ fixed point. 
In the presence of the symmetry-allowed Dirac mass that tunes the transition, the combination of these relevant perturbations leads to an inherently multicritical structure, and we argued the transition is generically first-order. By contrast, on tripartite lattices such as the $C_3$-symmetric breathing kagome, all monopoles can be symmetry-forbidden and a genuine, stable QED$_3$ critical point is possible.

A unifying message of our analysis is that the fate of ``trivial-to-trivial'' insulating transitions
depends sensitively on the UV to IR embedding of the lattice symmetries and not simply on band topology.
The observation that $N_f=4$ QED$_3$ on bipartite lattices admits a deformation to $N_f=2$ QCD$_3$ gives a robust route to identifying at least one monopole singlet under all microscopic symmetries, enforcing an instability of the QED$_3$ critical point in those settings. Conversely, when symmetry forbids all relevant monopoles, the emergent, critical gauge theory description is unfine-tuned.
We corroborated these conclusions with LSMOH anomaly matching for lattice bosons; every critical theory we discussed, including those that are dynamically unstable, matches the appropriate anomalies and is thus realizable in microscopic lattice models. 

Note that our methods can be straightforwardly extended to higher dimensions \cite{lin_prb_2018,luo_prb_2023}, such as for the pyrochlore \cite{ezawa_prl_2018} or cubic lattice obstructed atomic insulator.
However, due to the tendency toward deconfinement in higher dimensions, we expect a conformal critical point to be less common.
As QED$_4$ coupled to fermions is free in the IR and behaves like a pure gauge theory, one may need to employ parton constructions that involve a nonabelian gauge group.
In the setting of more general gauge groups, such as $SU(N_c)$, there is a conformal window of $N_f$ for which there exists a conformal fixed point (the Banks-Zak fixed point \cite{caswell_prl_1974,banks_nucl_1982}).
For $SU(2)$ fundamental fermions, we must have approximately $8\lesssim N_f \lesssim 11$, which is harder to realize in lattice models unless there are multiple degenerate bands.
However, these general theories could be an interesting avenue of exploration as $(3+1)d$ gauge theories realize a large class of ``unnecessary'' quantum critical points \cite{bi_prx_2019}, which may emerge as low energy descriptions in both fermionic and bosonic lattice insulators.

Moreover, in this paper we have only considered a particular type of transition: namely, those arising from Dirac fermions acquiring a mass.
More general transitions, such quadratic band touchings, will generically be described by some Lifshitz-type theory with gauge fields.
Looking forward, it would be fruitful to explore aspects of fractionalization and criticality in different settings such as in different dimensions and with more general internal symmetries.
While such critical points could be analyzed within a parton mean-field theory as we have done, we expect
many classes of critical points between atomic insulators to be beyond parton constructions and perhaps only describable intrinsically, such as in terms of anomalies and without reference to an explicit Lagrangian or Hamiltonian.

Lastly, whether a specific microscopic Hamiltonian realizes the QED$_3$ critical point or is preempted by a weakly first-order transition or intermediate phase is a dynamical question. 
Our analysis has identified universal properties and the conditions under which QED$_3$ criticality is allowed, and
as the models explored in this paper are relatively simple, there is hope that the unusual critical behavior, at least in the non-finetuned case of the breathing kagome lattice, can be probed in the future through both numerical simulations and laboratory experiments.

\section{Acknowledgments}
We thank Weicheng Ye, Zhengyan Darius Shi, Liujun Zou, and Chong Wang for useful discussions. YZ was supported by the National Science Foundation Graduate Research Fellowship
under Grant No. 2141064. TS was supported by NSF grant DMR-2206305.  

\appendix
\section{$C_4$-symmetric atomic insulator on the square: The BBH Model}
\label{app:bbh}
\subsection{The Fermionic BBH Model}
The BBH model 
\cite{bbh_sci_2017,bbh_prb_2017}
consists of four orbitals describing
spinless fermions on a square lattice.
There are dimerized hopping amplitudes corresponding to intra- and inter-site coupling.
Furthermore, $\pi$ flux is threaded through each plaquette as shown in Fig.~\ref{fig:bbh_lattice}.
The hopping Hamiltonian is given by,

\begin{align}
H_{BBH}&=\sum_{\vec{R}}\left[\lambda
(c_{1,\vec{R}}^\dagger c_{3,\vec{R}}+c_{2,\vec{R}}^\dagger c_{4,\vec{R}})\right.\nonumber\\
&\qquad+t(c_{1,\vec{R}}^\dagger c_{3,\vec{R}+\hat{x}}+c_{2,\vec{R}+\hat{x}}^\dagger c_{4,\vec{R}})\nonumber\\
&\qquad+\lambda(c_{1,\vec{R}}^\dagger c_{4,\vec{R}}-c_{2,\vec{R}}^\dagger c_{3,\vec{R}})
\nonumber\\
&\qquad\left.+t(c_{1,\vec{R}}^\dagger c_{4,\vec{R}+\hat{y}}-c_{2,\vec{R}+\hat{y}}^\dagger c_{3,\vec{R}})
\right]+h.c.,
\end{align}

We have written $c_{i,\vec{R}}$ as the fermion operator for the orbital/sublattice site $i$ within the unit cell $\vec{R}$.
The coupling $\lambda$ is the hopping amplitude within a unit cell, while $t$ is the intercell hopping amplitude to nearest neighbor unit cells.
Defining the Fourier transform $c_{i,\vec{R}} = \frac{1}{\sqrt{N}} \sum_{\vec{k}} e^{i\vec{k} \cdot R} c_{i,\vec{k}}$ and lattice vectors $\hat{x}=(1,0)$ and $\hat{y}=(0,1)$, we obtain the Bloch Hamiltonian,

\begin{align}
    h_{BBH}(\vec{k})&=(\lambda+t\cos k_1)\Gamma^4+t\sin k_1\Gamma^3\nonumber\\
    &+(\lambda+t\cos k_2)\Gamma^2+t\sin k_2\Gamma^1,
\end{align}
in terms of
the matrices $\Gamma^{i}$, where $\Gamma^4=\tau^1\otimes\tau^0$ and $\Gamma^{1,2,3}=-\tau^2\otimes\tau^{1,2,3}$ are anticommuting matrices $\left\{\Gamma^i,\Gamma^j\right\}=2\delta^{jl}$. We have written the Pauli matrices as $\tau^{i}$ and the identity matrix as $\tau^0$.
The energy bands are given by
\begin{equation}
        \epsilon(k)=\pm \sqrt{2}\sqrt{t^2+\lambda^2+\lambda t(\cos k_1+\cos k_2)},
\end{equation}
each of which is twofold degenerate.
\begin{figure*}[t]
\captionsetup{justification=raggedright}
{\includegraphics[width=2\columnwidth]{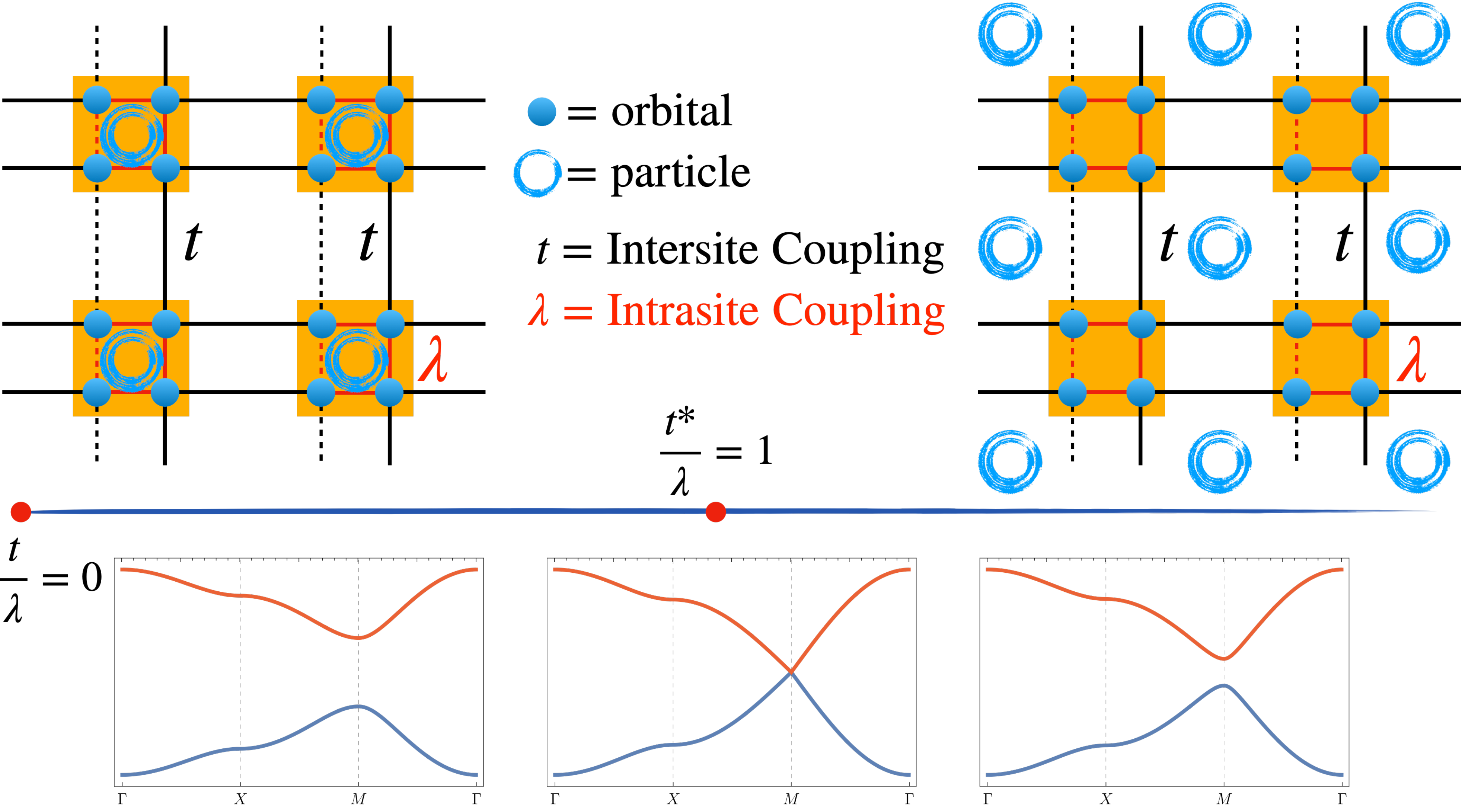}}
\caption{
Schematic phase diagram and dispersions at $t/\lambda$ is tuned across the critical point. Each of the two displayed bands are twofold degenerate.
We observe that the two phases are characterized by orbital centers localized either on the lattice sites or on the plaquettes between sites. At the
critical point, there is a touching of Dirac bands.}
\label{fig:bbh_phasediag}
\end{figure*}
Unless $t=\pm \lambda$, the spectrum is gapped, so at $\nu=2$ per unit cell the system is insulating.
At $t=\pm\lambda$, there is a phase transition as the spectrum becomes gapless at $\vec{M}=(\pi,\pi)$ [$\vec{\Gamma}=(0,0)$]. 
Upon tuning $t/\lambda$, one tunes between a phase whose Wannier orbitals are centered at the unit cells and a phase where the Wannier orbital centers are displaced by $(x,y)=(1/2,1/2)\pmod{1}$ relative to the unit cell centers.
These two phases cannot be connected by any adiabatic, $C_4$-symmetric deformation. The phase diagram, along with the dispersions of the Bloch bands, is shown in Fig.~\ref{fig:bbh_phasediag}.

There are two reflection symmetries $\mathcal{R}_{1,2}$, reversing the sign of $x$ and $y$, respectively, which act on the Bloch Hamiltonian as $\mathcal{R}_{1,2}h_{BBH}(\vec{k})\mathcal{R}_{1,2}^{\dagger}=h(\mathcal{R}_{1,2}\vec{k})$, where $\mathcal{R}_{1,2}=\tau^1\otimes \tau^{3,1}$.
The product of reflections also gives a $C_2$ symmetry with the operator $C_2=\mathcal{R}_1\mathcal{R}_2=-i\tau^0\otimes\tau^2$.
Lastly, we can define a $C_4$ symmetry     $(k_1,k_2)\rightarrow(k_2,-k_1)$ that acts on $h(\vec{k})$ as 
\begin{equation}
C_4=\begin{pmatrix}
    0&\tau^0\\-i\tau^2&0\end{pmatrix}=-\tau^2\otimes\tau^2e^{-i\frac{\pi}{2}\frac{\tau^0-\tau^3}{2}\otimes\tau^2}.
\end{equation}
In the presence of $\pi$ flux, the spatial symmetries as defined above are already projective.
Notably, the reflection symmetries do not commute,
\begin{equation}
    \left[\mathcal{R}_1,\mathcal{R}_2\right]=2i\tau^1\otimes\tau^2,\quad \left\{\mathcal{R}_1,\mathcal{R}_2\right\}=0.
\end{equation}
This non-commutation protects the twofold band degeneracy. Furthermore, one can check $C_4^4={C}_2^2=-1$.

Lastly, we note that the BBH model has time-reversal, chiral, and charge conjugation symmetries, acting as
\begin{align}
    \mathcal{T}h(\vec{k})\mathcal{T}^{-1}&=h(-\vec{k}),\quad \mathcal{T}=K,\\
    \Pi h(\vec{k})\Pi^{-1}&=-h(\vec{k}),\quad \Pi=\Gamma^0=\tau^3\otimes\tau^0,\\
    \mathcal{C} h(\vec{k})\mathcal{C}^{-1}&=-h(\vec{k}),\quad \mathcal{C}=\Pi=\Gamma^0.\label{eq:chargeconj_bbh}
\end{align}
The effective Hamiltonian at the Dirac point $\vec{M}=(\pi,\pi)$ near the critical point $t=\lambda$ is 
\begin{align}
    &H_{BBH}(\vec{k}=\vec{M}+\vec{q})\approx \nonumber\\&\psi_{\vec{q}}^{\dagger}\left[-t q_1\Gamma^3-t q_2\Gamma^1+(t-\lambda)(\Gamma^4+\Gamma^2)\right]\psi_{\vec{q}}.
\end{align}
To simplify the above, we will define the spinor,
\begin{equation}
    \begin{pmatrix}
    \Psi_{1}(\vec{q})
    \\\Psi_{2}(\vec{q})
\end{pmatrix}\equiv e^{-i\pi\tau^0\otimes\tau^1/4}e^{-i\pi\tau^2\otimes\tau^2/4}\psi_{\vec{q}},
\end{equation}
from which we can write (in units of $t=1$)
\begin{equation}
\label{eq:dirac_bbh}        H_{BBH,crit}=\Psi_{\alpha}^{\dagger}(\vec{q})\left[q_1\tau^1+q_2\tau^2+
        m(\mu^2+\mu^3)\tau^3\right]\Psi_{\alpha}(\vec{q}),
\end{equation}
where we have denoted $\mu^i$ to act on the "valley" index $\alpha$, $\tau^i$ to act in the two-dimensional Dirac spinor space, and omitted the tensor product $\otimes$ for ease of notation.
Therefore, the phase transition is described by $2$ Dirac fermions, with mass $m\equiv 1-\lambda$.
Note that in the low energy theory, $C_4$ acts as $\frac{i}{\sqrt{2}}(\mu^2+\mu^3)e^{i\frac{\pi}{4}\mu^0\tau^3}$, while the reflections act as $\mathcal{R}_{1,2}=-\mp \mu^1\tau^{2,1}$ and time reversal acts as $\mathcal{T}=-i\mu^2\tau^2$.
The particle-hole symmetry defined in Eq.~\eqref{eq:chargeconj_bbh} acts as $\mathcal{C}=i\mu^3\tau^1$.

Under these symmetries, the only allowed mass term is of the form $(\mu^2+\mu^3)\tau^3$, so the Dirac theory is indeed a quantum critical point between two different trivial insulator phases.
This theory has been extensively studied due to its emergence in graphene, and short-range interactions are irrelevant at the critical point.
Furthermore, transitions in fermionic obstructed atomic insulators
arising from the annihilation and reappearance of Dirac cones have been analyzed in \cite{radha_prb_2010}.

\subsection{The Bosonic BBH model}
To analyze the bosonic BBH model, 
we will take a single species of hardcore boson $b$ hopping on the square lattice, and fractionalize it in terms of fermionic partons $d_{1,2}$,
\begin{equation}
    b=d_1^{\dagger}d_2.
\end{equation}
This rewriting reproduces the physical Hilbert space given we impose the constraint $d_1^{\dagger}d_1+d_2^{\dagger}d_2=1$ at each site/orbital and set the filling $\nu_b=\nu_{d_2}$.
We will only impose the constraints on average to arrive at a mean field Hamiltonian
\begin{equation}
\label{eq:bbh_parton}
H_{MF}(b^{\dagger},b)=H_{BBH}(d_1^{\dagger},d_1)+H_{BBH}(d_2^{\dagger},d_2),
\end{equation}
where we have put each fermionic parton at half filling in the BBH model. 
We do not attempt to specify an explicitly form of the original bosonic Hamiltonian;
as long as it respects all the UV symmetries, it can take a general form, though we assume there is some parton mean field decoupling 
such that the mean field state in Eq.~\eqref{eq:bbh_parton} is realized.

Before proceeding further, let us explicitly outline how the spatial symmetries act projectively on the fermions.
For the fermion $d_{\alpha,i,\vec{R}}$ of species $\alpha$ on orbital site $i$, we have (omitting the index $\alpha$ for ease of notation)
\begin{align}
    {C}_4&\;:\; d_{[1,2,4],\vec{R}}\rightarrow d_{[3,4,1],C_4\vec{R}},\quad d_{3,\vec{R}}\rightarrow -d_{2,C_4\vec{R}},\nonumber\\
    {C}_2
    &\;:\; d_{[2,4],\vec{R}}\rightarrow d_{[1,3],C_2\vec{R}},\quad d_{[1,3],\vec{R}}\rightarrow -d_{[2,4],C_2\vec{R}},\nonumber\\
    \mathcal{R}_x
    &\;:\; d_{[1,3],\vec{R}}\rightarrow d_{[3,1],R_1\vec{R}},\quad d_{[2,4],\vec{R}}\rightarrow -d_{[4,2],C_1\vec{R}},\nonumber\\    
        \mathcal{R}_y
    &\;:\;d_{[1,2,3,4],\vec{R}}\rightarrow d_{[4,3,2,1],R_1\vec{R}}.
\end{align}
The time reversal and charge conjugation symmetries can be chosen to act as $\mathcal{T}(d_{1},d_{2})=(-d_2,d_1)$ and \begin{equation}
    C\;:\;d_{[1,2],\vec{R}}\rightarrow d_{[1,2],\vec{R}}^{\dagger},\quad d_{[3,4],\vec{R}}\rightarrow -d_{[3,4],\vec{R}}^{\dagger}.
\end{equation}
We will also
maintain full $SU(2)$ pseudospin symmetry throughout, in addition to  $U(1)_b$ number conservation symmetry of the hardcore bosons, which rotates $d_{1,2}\rightarrow e^{i\sigma^3}d_{1,2}$.
We observe that the time reversal symmetry is what constrains both fermionic partons to simultaneously go critical.
Note that within the decomposition $b=d_1^{\dagger}d_2$, there is an inherent $SU(2)_g$ gauge symmetry $(d_1^{\dagger},d_2)\rightarrow U (d_1^{\dagger},d_2)$.
This is the same framework used in parton constructions of spin liquid states \cite{wen_prb_2002}.
The mean field Hamiltonian $H_{MF}$ actually preserves the full $SU(2)_g$ gauge symmetry inherent in the parton description.
One way to see this is to rewrite
\begin{equation}
    H_{MF}=\sum_{i=1,2}\sum_{\langle\vec{r},\vec{r}'\rangle} \begin{pmatrix}
        d^{\dagger}_{1,\vec{r}}&d^{}_{2,\vec{r}}
    \end{pmatrix}H_{\vec{r},\vec{r}'}\begin{pmatrix}
        d_{1,\vec{r}}\\d^{\dagger}_{2,\vec{r}}
    \end{pmatrix}.
\end{equation}
As for nearest neighbors $\vec{r}$ and $\vec{r}'$,
\begin{equation}
    H_{\vec{r},\vec{r}'} \propto\begin{pmatrix}
        1&\\&-1
    \end{pmatrix},
\end{equation}
we see that any product over $H$'s over a lattice Wilson loop $\mathcal{C}$ will contain an even number of bonds and therefore have trivial $SU(2)_g$ flux.
\begin{equation}
\prod_{\mathcal{C}}H_{\vec{r},\vec{r}'}=H_{\vec{r}_0,\vec{r}_1}H_{\vec{r}_1,\vec{r}_2}\cdots H_{\vec{r}_n,\vec{r}_0}\propto \mathbb{I}.    \end{equation}
Consequently, the invariant gauge group is the full $SU(2)_g$, and at the critical point, the effective theory that emerges will be a form of QCD$_3$.
We would like to Higgs the $SU(2)_g$ to $U(1)$, which can be done by same sublattice or imaginary hopping terms such
\begin{align} 
\label{eq:bbh_higgs}&it_2(d_{1,\vec{R}}^{\dagger}d_{3,\vec{R}+\hat{x}}-d_{1,\vec{R}}^{\dagger}d_{4,\vec{R}+\hat{y}}+
d_{2,\vec{R}}^{\dagger}d_{3,\vec{R}-\hat{y}}+d_{2,\vec{R}}^{\dagger}d_{4,\vec{R}-\hat{x}})\nonumber\\
&+i\lambda_2(d_{1,\vec{R}}^{\dagger}d_{3,\vec{R}}-d_{1,\vec{R}}^{\dagger}d_{4,\vec{R}}+
d_{2,\vec{R}}^{\dagger}d_{3,\vec{R}}+d_{2,\vec{R}}^{\dagger}d_{4,\vec{R}})+h.c.
\end{align}
with $t_2,\lambda_2\propto t,\lambda$ and taking the same coupling for each pseudospin species.
Note that these couplings break the particle-hole symmetry $\mathcal{C}$, in addition to $\mathcal{T}$ and the reflection symmetries.
Instead, in the presence of these imaginary hoppings, the actions of $\mathcal{T}$ and the reflection symmetries should be combined with $\mathcal{C}$, under which the model is still invariant.
The imaginary hoppings essentially add a staggered flux to the plaquettes of the uniform $\pi$-flux BBH model.
In the low energy Dirac theory (focusing on a single species of fermionic parton), at the critical point $t=\lambda$ (so $t_2=\lambda_2$), the particle-hole breaking terms lead to anisotropic
Dirac velocities $\Psi^{\dagger}(q_1\mu^1\tau^2+q_2\mu^1\tau^1)\Psi$.
Then, at the critical point, the $SU(2)_g$ gauge field is Higgsed to $U(1)$, and 
we can write the total Lagrangian as (performing a particle-hole transformation on $d_1$ so $d_{1,2}$ carry equal charge under the emergent gauge field),
\begin{equation}
\label{eq:qed3_lag}\mathcal{L}_E=\sum_{a=1}^4\ol{\Psi}_a(-i\gamma^{\mu}(\del_{\mu}+ia_{\mu}))\Psi_a+\frac{1}{4e^2}f_{\mu\nu}^2+\delta\mathcal{L}_E,
\end{equation}
where $a$ goes from $1$ to $N_f=4$, labeling the $2$ Dirac valleys, each with $2$ flavors of fermionic partons.
The Dirac fermions are minimally coupled to an emergent $U(1)$ gauge field $a$ with curvature $f_{\mu\nu}=\del_{\mu}a_{\nu}-\del_{\nu}a_{\mu}$. 
We have defined
$\gamma^{\mu}=(\tau^3,-\tau^2,\tau^1)$ and $\ol{\Psi}=i\Psi^{\dagger}\tau^3$.
The term $\delta\mathcal{L}_E$ contains all symmetry allowed perturbations, including velocity anisotropies. 
Eq.~\eqref{eq:qed3_lag} is exactly $N_f=4$ QED$_3$, which has emerged at this critical point.
The general background for this critical theory and its relevant monopole operators is described in the main text. To analyze whether the QED$_3$ critical point is stable, we will analyze if there are any symmetry allowed-monopoles.

\subsection{Monopole quantum numbers}
For convenience, we will calculate the symmetry properties of the monopole operators in the original BBH model with no particle-hole symmetry breaking perturbations as such perturbations do not affect the monopole quantum numbers.
We adopt the same basis of monopole operators that was outlined in the main text,
\begin{align}
    \phi^{\dagger}_{1,2,3}&=f^{\dagger}\left(i\mu^2\mu^{1,2,3}\otimes i\sigma^2\right)f^{\dagger}\mathcal{M}_{bare},\\
    \phi^{\dagger}_{4,5,6}&=if^{\dagger}\left(i\mu^{2}\otimes i\sigma^2\sigma^{1,2,3}\right)f^{\dagger}\mathcal{M}_{bare}
\end{align}

Recall the low energy theory, Eq.~\eqref{eq:qed3_lag}, is that of $N_f=4$ QED$_3$ on  the square lattice, with all Dirac nodes located at momentum $\vec{M}=(\pi,\pi)$.
In the Dirac basis, the physical symmetries act as
\begin{align}
    T_{1,2}\;:\;&\Psi\rightarrow -\Psi,\\
    \mathcal{R}_{x}\;:\;&\Psi\rightarrow \tau^2\mu^1\Psi,\\
    C_4\;:\;&\Psi\rightarrow \frac{i}{\sqrt{2}}e^{i\frac{\pi}{4}\tau^3}(\mu^2+\mu^3)\Psi,\\
    \mathcal{T}\;:\;&\Psi\rightarrow -i\tau^2\mu^2\sigma^2\Psi.
\end{align}
Furthermore, we will define a charge conjugation $\mathcal{C}$ symmetry in the parton model that acts $\mathcal{C}$ in Eq.~\eqref{eq:chargeconj_bbh} in addition to a spin flip on the partons,
\begin{equation}
\mathcal{C}\;:\;\Psi\rightarrow i\tau^1\mu^3\sigma^2\Psi^*.
\end{equation}
Up to a $U(1)_{top}$ phase, the above symmetries determine the symmetry transformations of the monopoles.

Now to fully constrain the monopole transformations, especially the Berry phases of the UV lattice symmetry transformations, we will again borrow from \cite{song_prx_2020}. 
Note that due to the particle-hole symmetry inherent in the bipartite square lattice,
the hopping Hamiltonian can be adiabatically tuned to a point with an enhanced $SU(2)_g$ gauge symmetry, as in the case of taking the diagonal hopping terms $\lambda_2\rightarrow 0$.
A particle-hole symmetry is then restored at this point.
For example, for a site $i_a$ on the $a$ sublattice, there is a symmetry
\begin{equation}
d_{\alpha,i_a}^{}\rightarrow (-1)^{a}i\sigma^2_{\alpha\beta}d^{\dagger}_{\beta,i_a},
\end{equation}
where $(d_{\alpha,i_a},i\sigma^2_{\alpha\beta}d^{\dagger}_{\beta,i_a})$ forms an $SU(2)_g$ fundamental.
The resulting critical point is then described $N_f=2$ QCD$_3$. 
At low energies, QED$_3$ can recovered if $SU(2)_g$ is Higgsed.

Equivalently, we can view the Higgsed QCD$_3$ as a mid-IR theory in which $SU(2)_g$ has been Higgsed down to $U(1)$, while the particle-hole symmetry still survives as a global $\mathbb{Z}_2$ symmetry.
In either case, there turns out to be a unique embedding of implementation of the lattice UV symmetries into the QCD$_3$, such that the QED$_3$ theory that will contain a trivial (symmetry-allowed) spin-singlet monopole.
Specifically, the $SO(5)$ flavor symmetry of the $N_f=2$ QCD$_3$ theory will embed into the $SO(6)\times U(1)_{top}/\mathbb{Z}_2$ of QED$_3$ in such a way that five of the six monopoles transform as an $SO(5)$ vector, while the remaining singlet monopole will be trivial under all symmetries.
Because it is a singlet in the mid-IR theory, this singlet monopole will remain a singlet in the IR QED$_3$ theory, as the process of flowing the IR by adding charge conjugation symmetry breaking and other irrelevant perturbations will not change any monopole quantum numbers. We have thus shown the presence of one trivial, symmetry-allowed monopole operator.

From another point of view, the map of symmetry groups
\begin{equation}
    G_{UV}\rightarrow G_{QED_3}
\end{equation}
must factor through the mid-IR QCD$_3$ theory, such that we have a map,
\begin{equation}
    G_{UV}\rightarrow G_{QCD_3}\rightarrow G_{QED_3}.
\end{equation}
Then the embedding $G_{QCD_3}\rightarrow G_{QED_3}$ guarantees a trivial monopole, from which we conclude $    G_{UV}\rightarrow G_{QED_3}$ must also lead to a trivial monopole.
This feature is present whenever the $U(1)$ gauge theory that can be adiabatically tuned to have $SU(2)_g$ gauge symmetry, as in the case generically for parton models on bipartite lattices. 

We can now apply this knowledge. To begin, we see that all the monopoles are trivial under translation.
Under $C_4$, we have $\phi^{\dagger}\rightarrow e^{i\theta_{C_4}}O_{C_4}\phi^{\dagger}$, where
\begin{equation}
    O_{C_4}=\begin{pmatrix}
    -1&&&&&\\
    &&1&&&\\
    &1&&&&\\
    &&&1&&\\
    &&&&1&\\
    &&&&&1
\end{pmatrix}.
\end{equation}
However, from the fact that there exists a singlet monopole, we know that the $U(1)_{top}$ phase should be chosen so that $O_{C_4}$ takes the form \begin{equation}
O_{C_4}\begin{pmatrix}
    1&\\&U
\end{pmatrix},
\end{equation}
for $U\in SO(5)$.
Therefore, $\theta_{C_4} =\pi$.
The time reversal $\mathcal{T}$ acts exactly the bare time reversal $\mathcal{T}_{IR}$ defined in Eq.~\eqref{eq:baresym_monopole}.
Note we have used here that the $SO(3)_f$
pseudospin rotation symmetry must be part of the
$SO(5)$, so that the $SO(5)$ singlet monopole must be
a spin singlet (so one of $\phi_{1,2,3}$).

For the symmetries of charge conjugation and reflection,
we observe that the operator $\mathcal{C}\mathcal{R}_x$ can be viewed as a combination of $\mathcal{C}_{IR}\mathcal{R}_{x,IR}$ and a Lorentz rotation that does not affect the monopoles, which are Lorentz scalars. The overall phase of
$\mathcal{C}\mathcal{R}_x$ can be fixed by noting that $(\mathcal{C}\mathcal{R}_x)^2=1$ on the fermions. From the arguments in \cite{song_prx_2020}, $(\mathcal{C}\mathcal{R}_x)^2=1$ is another manifestation of the nontriviality of the spin Hall insulator, in which the spin triplet monopoles must be odd under $\mathcal{CR}_x$. Therefore, the (spin triplet) monopoles in this system must also transform the same way, which uniquely determines the combined action $\mathcal{C}\mathcal{R}_x$.
Note that one of $\mathcal{C}$ and $\mathcal{R}_x$ can be defined with an arbitrary overall $U(1)_{top}$ phase, as this amounts to shifting all the monopoles by an overall phase.
However, their relative actions are physically relevant and determined by the arguments above.
The symmetry transformations are gathered in Table~\ref{table:monopole_fund_bbh}.

\begin{table}[h]
\captionsetup{justification=raggedright}
\begin{center}
\begin{tabular}{|c|c|c|c|c|c|}
\hline
Monopole                 & $T_{1,2}$              & $C_4$              & $\mathcal{R}_x$             & $\mathcal{C}$          & $\mathcal{T}$             \\ \hline
$\phi_1^{\dagger}$       & $\phi_1^{\dagger}$        & $\phi_1^{\dagger}$       & $\phi_1^{}$      & $\phi_1^{}$      & $\phi_1^{}$        \\ 
{$\phi_2^{\dagger}$}       & $\phi_2^{\dagger}$       & $-\phi_3^{\dagger}$       & $-\phi_2^{}$      & $-\phi_2^{}$      & $\phi_2^{}$       \\ 
$\phi_3^{\dagger}$       & $\phi_3^{\dagger}$       & $-\phi_2^{\dagger}$        & $-\phi_3^{}$       & $-\phi_3^{}$       & $\phi_3^{}$        \\ 
$\phi_{4/5/6}^{\dagger}$ & $\phi_{4/5/6}^{\dagger}$ & $-\phi_{4/5/6}^{\dagger}$ & $\phi_{4/5/6}^{}$ & $-\phi_{4/5/6}^{}$ & $-\phi_{4/5/6}^{}$ \\ \hline
\end{tabular}
\end{center}
\caption{The transformation of the single-charge monopoles under the UV symmetries in the square lattice critical point.}
\label{table:monopole_fund_bbh}
\end{table}

We see straightforwardly that the real part of $\phi_1$, $\Re[\phi_1]\sim \phi_1^{\dagger}+\phi$ is a trivial monopole.
However, we note one surprise: $\Im[\phi_2-\phi_3]\sim i(\phi_2^{\dagger}-\phi_3^{\dagger}-\phi_2+\phi_3)$is also symmetry allowed!
More generally, the insertion of either of these operators into the Lagrangian is allowed by symmetry, and
as these operators are strongly relevant at the critical point,
we then do not expect the $N_f=4$ QED$_3$ CFT to be the ultimate fate of the critical point.
Instead, the presence of these two monopoles, $\langle\Re[\phi_1]\rangle,\langle\Im[\phi_2-\phi_3]\rangle\ne 0$, will destabilize the QED$_3$ theory and destroy the conformal fixed point, leading to what is most likely a first-order transition.

\subsubsection{Multicriticality}
Note the resulting phase, between two different atomic insulators on the square lattice, is multicritical, as it requires the fine tuning of three independent (RG relevant) parameters. 
Recall that to reach the QED$_3$ critical point, one must already tune the mass $i\ol{\Psi}(\mu_2+\mu_3)\Psi$ from Eq.~\eqref{eq:dirac_bbh} to zero.
This mass term is equivalent to the monopole-antimonopole pair, 
\begin{equation}
 \Im[\phi_1^{\dagger}(\phi_2-\phi_3)].
\end{equation}
There are also other allowed monopole-antimonopole terms such as $\phi_1^{\dagger}\phi_1$, and these terms have scaling dimension $\sim 2.38$ from large $N_f$.
However, as mentioned earlier, we will assume that they are irrelevant as is done in numerical calculations, due to the fact that any lattice realization of QED$_3$ will contain singlets of this form \cite{he_scipost_2022,zhang_scipost_2025,zhang2025unnecessaryquantumcriticalitysu3}.

From the case of the BBH model, we observe that in all transitions between bosonic atomic insulators in which the critical point is accessed through tuning a fermion billinear mass $i\overline{\Psi}M\Psi$ to zero, the existence of a single trivial monopole $\phi_{triv}$ may imply the existence of an additional one.
This is because in addition to the trivial monopole $\phi_{triv}$, the 
tuning mass $i\overline{\Psi}M\Psi$ is also an allowed relevant operator and is equivalent to a monopole-antimonopole pair $\sim\Im[\phi_{M_1}^{\dagger}\phi_{M_2}]$ for some $\{M_{1},M_2\}$.
These operators can be identified because a zero flux monopole-antimonopole composite transforms as the antisymmetric two indexed tensor representation of $SO(6)$ ($\mathbf{15}\subset \mathbf{6}\otimes\mathbf{6}$), which is exactly the adjoint of $SU(4)$.
Then, one can see that in the case that $\phi_{triv}$ coincides with one of $\phi_{M_{1,2}}$, then $\phi_{M_{2,1}}$ will also be symmetry allowed.
Therefore, in cases in which there is one trivial monopole (as is generically the
case for the square and honeycomb lattices), we would expect such critical points to be highly multicritical and unstable as in the case we just examined.

The resulting multicritical point will presumably be driven to strong coupling by the two monopole operators, and will have a reduced symmetry 
\begin{equation}
    \dfrac{SO(6)\times U(1)_{top}}{\mathbb{Z}_2}\rightarrow SO(4)\times SO(2),
\end{equation}
where the $SO(2)$ rotates among the two singlet monopoles.
The remaining state hosts an $SO(4)$ symmetry anomaly, so the theory cannot flow to an $SO(4)$ symmetric gapped state.
To see this, one can imagine adding the monopoles one at a time. 
Adding the first monopole breaks $SO(6)\rightarrow SO(5)$, leading to the anomaly for $N_f=2$ QCD$_3$, with bulk partition function on a 4-manifold $X_4$
\begin{equation}
    Z[A^{SO(5)},X_4]=\exp\left(\pi i\int_{X_4}w_4[A^{SO(5)}]\right),
\end{equation}
for $w_4[A^{SO(5)}]\in H^4(X_4,\mathbb{Z}_2)$, corresponding to a discrete  $\theta$-angle for the background $SO(5)$ gauge field $A^{SO(5)}$.
Proliferating the second trivial monopole will restrict our symmetry to an $SO(4)$ subgroup of $SO(5)$.
Decomposing $SO(4)=(SU(2)_L\times SU(2)_R)/\mathbb{Z}_2$, the anomaly becomes \cite{wang_prx_2017}
\begin{widetext}
  \begin{equation}
    Z[A^{SU(2)_L},A^{SU(2)_R},X_4]=\exp\left(\frac{i}{2}CS_{SU(2),X_4}[A^{SU(2)_L}]-\frac{i}{2}CS_{SU(2),X_4}[A^{SU(2)_R}]\right),
\end{equation}  
\end{widetext}
where $CS_{SU(2),X_4}[A]=\frac{1}{4\pi}\int_{X_4}\operatorname{tr}_{fund}F\wedge F$ is the extension of the Chern-Simons term into $X_4$.
This anomaly corresponds to opposite discrete $\theta$-angles for $SU(2)_L$ and $SU(2)_R\subset SO(4)$, and with time reversal, the $\theta$ angles do not flow.
Most importantly, the resulting theory cannot be a trivial, symmetric gapped phase.
There is a possibility that in the
presence of two singlet monopoles, the critical point flows to a conformal fixed point distinct from QED$_3$.
Note that despite the 't Hooft anomaly, the state enjoys symmetry-enforced gaplessness \cite{wang_prb_2014,wang_prb_2016,sodemann_prb_2016,wang_prx_2017,cordova_arxiv_2019,cordova_prd_2020}, in which the anomaly precludes even a gapped symmetric topologically ordered state.
The argument is similar to the one in \cite{dumitrescu_arxiv_2024}.
Because $SU(2)$ is simply-connected, it cannot have a non-trivial action on a $(2+1)d$ TQFT; that is, $SU(2)$ cannot carry any non-trivial anomaly.
However, the $\theta$ terms imply a non-trivial anomaly for an $SU(2)$ subgroup of $SO(4)$.
Instead, a topologically ordered state must break the $SO(4)$ or time reversal (in which case the $\theta$ can flow to be trivial).

Therefore, as in the case posited for a single monopole \cite{zhang_scipost_2025,zhang2025unnecessaryquantumcriticalitysu3}, it is more plausible that the final fate of the theory could be some ordered state that spontaneously breaks the remaining $SO(4)$ symmetry. 
In either case, the highly multicritical nature of the critical point leads us to conclude the transition we have constructed is likely first-order.

We observe that in the above picture, the two insulating phases of the BBH model near the critical point correspond to turning on a suitable $SU(4)$ adjoint mass for $\psi$ while not proliferating any of the two singlet monopoles.
This gaps out the fermion matter fields, leaving a pure gauge theory in the IR.
Upon turning on the mass, the degeneracy of the six monopoles will be lifted, and there will be one unique gapless monopole in the IR, specifically $\phi_1\pm i\frac{\phi_2-\phi_3}{\sqrt{2}}$.
This monopole will spontaneously proliferate as it is also a singlet under the UV symmetries,
leading to confinement of the resulting gauge theory.
The resulting state (in addition to the discrete symmetries of $\mathcal{C}_{IR}$, $\mathcal{T}_{IR}$, and $\mathcal{R}_{IR}$)
will have an $SO(4)\times SO(2)/\mathbb{Z}_2$ symmetry, where $SO(2)$ acts as a combination of $U(1)_{top}$ and rotating among the singlet monopoles while $SO(4)$ rotates the other four and $\mathbb{Z}_2$ acts as before.
A simple ordered state is allowed, and consequently, there is no anomaly even in the IR. However, we still expect the IR symmetry to completely break the global symmetry to the symmetry of the UV square lattice space group. 
This is because in the presence of the adjoint mass, the UV singlet terms allowed in $\delta{\mathcal{L}}_E$ that were irrelevant at the critical point (such as higher strength monopoles) will become relevant and will completely break the IR continuous vector symmetries into discrete subgroups corresponding to the UV lattice symmetries.
We remark that the activation of an adjoint mass in QED$_3$ is also a possible mechanism by which conventional Landau symmetry-breaking orders can originate from the a Dirac spin liquid phase or critical point, also described by QED$_3$ \cite{song_nat_2019}. However, our case is distinct in that the adjoint mass is not a mechanism for symmetry breaking as it is already a singlet under all UV symmetries.

\section{$C_6$-symmetric atomic insulator on the breathing honeycomb lattice}
\label{app:c6}
\subsection{Fermionic}
We begin with a hopping model on the breathing honeycomb lattice with real couplings as shown in Fig.~\ref{fig:tri_lattices}(a).
The spectrum is gapped at half-filling, with two Dirac cones at $\vec{k}=\vec{\Gamma}$. For $\lambda>t$, the phase hosts Wannier orbitals at the hexagonal plaquette centers, while for $t>\lambda$, the Wannier centers are located at the hexagonal plaquette edges.
At quantum critical point $t=\lambda$, the spectrum becomes gapless.
The model's symmetries include translation $T_{1,2}$, reflection $\mathcal{R}_x$, time reversal $\mathcal{T}$, particle-hole symmetry $\mathcal{C}$ (from the bipartite nature of the lattice), and $C_6$ rotation about a plaquette.
Near the critical point, we obtain an effective Hamiltonian near the Dirac points
\begin{equation}
        H_{eff}=\Psi_{\alpha}^{\dagger}(\vec{q})\left[\tilde{q}_1\tau^1+\tilde{q}_2\tau^2+
        m(\sqrt{3}\mu^2-\mu^3)\tau^3\right]\Psi_{\alpha}(\vec{q}),
\end{equation}
where $\tilde{q}_1=-q_1+\sqrt{3}q_2$ and $\tilde{q}_2=-\sqrt{3}q_1-q_2$.
As previously, $\tau$ acts on the spinor/Lorentz index and $\mu$ acts on the valley index $\alpha$.
We have defined the mass $m=\frac{1-\lambda}{2}$ in units of $t=1$.
By the microscopic lattice symmetries, no other mass terms are allowed.
Therefore, the Dirac theory is a non-finetuned critical point between the two diferent fermionic trivial insulating phases, as before.
\subsection{Bosonic}
\begin{figure*}[t]
\captionsetup{justification=raggedright}
\centering
\subcaptionbox{}
{\includegraphics[width=2\columnwidth]{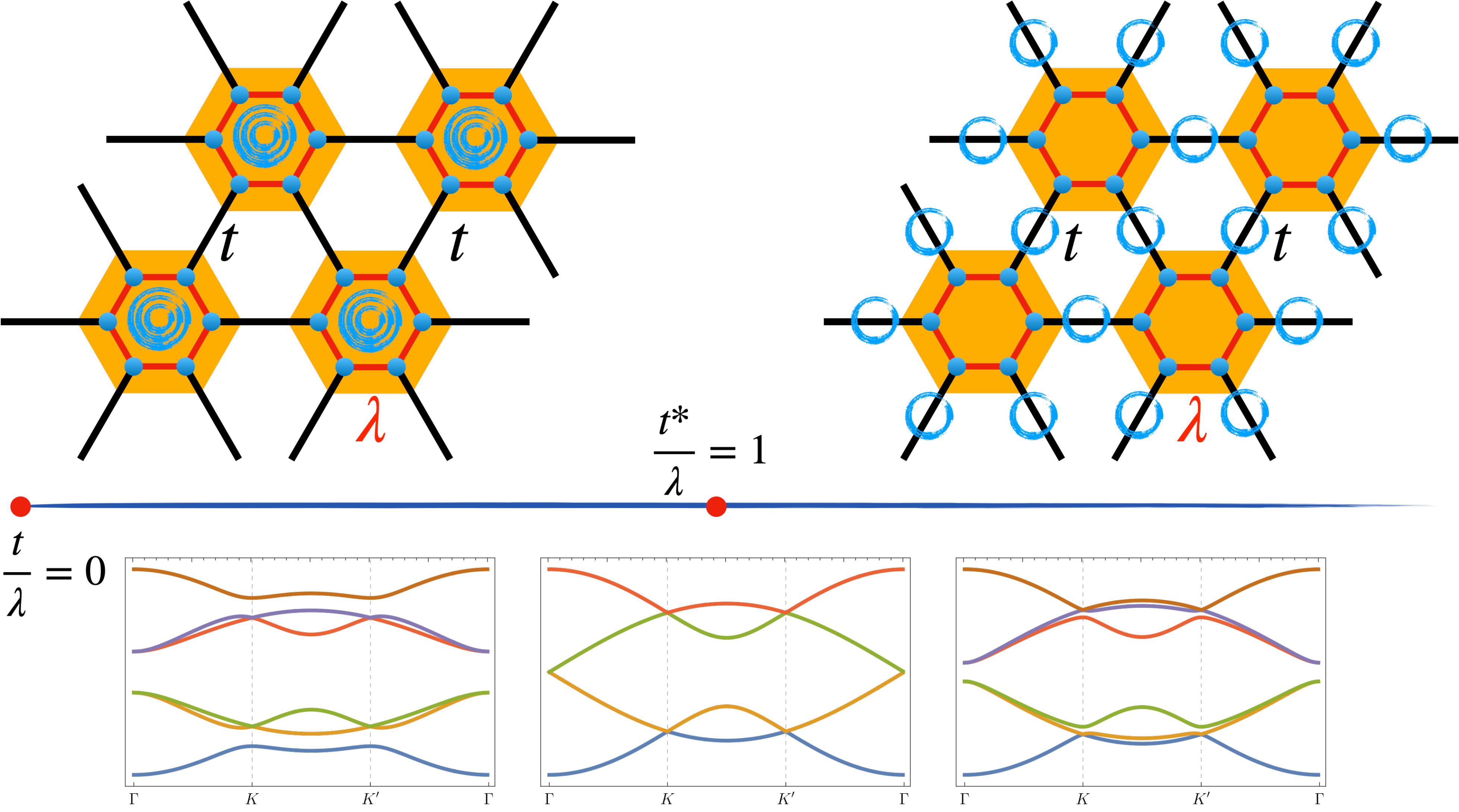}}
\caption{Schematic phase diagram and dispersions as $t/\lambda$ is tuned across the critical point. At the critical point,
there are Dirac cones at $\vec{\Gamma}$.
}
\label{fig:hex_phasediag}
\end{figure*}

In the bosonic case, we repeat the steps done before, fractionalizing the hardcore boson on the lattice into fermionic partons $b=d_1^{\dagger}d_2$, having the partons each realize the Hamiltonian in Fig.~\ref{fig:hex_phasediag}, and half-filling each fermionic parton.

The resulting theory is again that of $N_f=4$ QED$_3$,
\begin{equation}
\mathcal{L}_E=\sum_{N_f=4}\ol{\Psi}(-i\gamma^{\mu}(\del_{\mu}+ia_{\mu}))\Psi+\frac{1}{4e^2}f_{\mu\nu}^2+\delta \mathcal{L}_E,
\end{equation}
where $\delta \mathcal{L}_E$ contains irrelevant contributions and velocity anisotropy terms that Higgs the $SU(2)_g$ gauge symmetry to $d_{1,2}$ down to $U(1)$.
Repeating the steps done for the square lattice, we will project the action of the lattice symmetries onto the Dirac fermions to find
\begin{align}
    T_{1,2}\;:\;&\Psi\rightarrow \Psi,\\
    \mathcal{R}_{x}\;:\;&\Psi\rightarrow \frac{1}{2}(-\tau^2\mu^1+\sqrt{3}\tau^1\mu^1)\Psi,\\
    C_6\;:\;&\Psi\rightarrow \frac{1}{4}(i\sqrt{3}\tau^0\mu^3-\tau^3\mu^3-i3\tau^0\mu^2+\sqrt{3}\tau^3\mu^2)\Psi,\\
    \mathcal{T}\;:\;&\Psi\rightarrow -i\tau^2\mu^2\sigma^2\Psi,\\
    \mathcal{C}\;:\;&\Psi\rightarrow i\tau^1\mu^3\sigma^2\Psi^*.
\end{align}
Importantly, the breathing honeycomb lattice is also bipartite and will therefore host at least one trivial monopole.
Using the same methods as earlier, we find the transformation of all monopole operators, tabulated in Table~\ref{table:monopole_fund_c6}.
We comment that the Berry phase for $C_6$ is given by $\theta_{C_6}=\pi$, and the action of charge conjugation and reflection can again be determined by seeing that $(\mathcal{CR}_x)^2=1$, which means the spin triplet monopoles will be odd under $\mathcal{CR}_x$.

\begin{table}[h]
\captionsetup{justification=raggedright}
\begin{center}
\begin{tabular}{|c|c|c|c|c|c|}
\hline
Monopole                 & $T_{1,2}$              & $C_6$              & $\mathcal{R}_x$             & $\mathcal{C}$          & $\mathcal{T}$             \\ \hline
$\phi_1^{\dagger}$       & $\phi_1^{\dagger}$        & $\phi_1^{\dagger}$       & $\phi_1^{}$      & $\phi_1^{}$      & $\phi_1^{}$        \\ 
{$\phi_2^{\dagger}$}       & $\phi_2^{\dagger}$       & $\dfrac{-\phi_2^{\dagger}+\sqrt{3}\phi_3^{\dagger}}{2}$       & $-\phi_2^{}$      & $-\phi_2^{}$      & $\phi_2^{}$       \\ 
$\phi_3^{\dagger}$       & $\phi_3^{\dagger}$       & $\dfrac{\sqrt{3}\phi_2^{\dagger}+\phi_3^{\dagger}}{2}$        & $-\phi_3^{}$       & $-\phi_3^{}$       & $\phi_3^{}$        \\ 
$\phi_{4/5/6}^{\dagger}$ & $\phi_{4/5/6}^{\dagger}$ & $-\phi_{4/5/6}^{\dagger}$ & $\phi_{4/5/6}^{}$ & $-\phi_{4/5/6}^{}$ & $-\phi_{4/5/6}^{}$ \\ \hline
\end{tabular}
\end{center}
\caption{The transformation of the single-charge monopoles under the UV symmetries in the breathing honeycomb lattice critical point.}
\label{table:monopole_fund_c6}
\end{table}
\subsection{Multicriticality}
As found in the case of the square lattice, there are two trivial monopoles: $\Re[\phi_1]$ and $\Im[\sqrt{3}\phi_2/3+\phi_3]$.
From the bipartite nature of the lattice, there is an adiabatic deformation to an effective QCD$_3$ theory that necessitates the existence of the trivial monopole $\Re[\phi_1]$. 
Then, the combination of $\Re[\phi_1]$ and the tuning mass leads to the existence of a second trivial monopole $\Im[\sqrt{3}\phi_2/3+\phi_3]$.
Again, such a structure of singlet monopoles is expected as the tuning mass is exactly the composite monopole operator $\Im\left[\phi_1(\sqrt{3}\phi_2/3+\phi_3)\right]$.

Compared with the BBH square lattice insulator, an essentially identical analysis applies here, from which we conclude that the quantum critical point is actually multicritical,
leading to what is likely a first-order transition between the two atomic insulating phases.
As before, aside from some more exotic possibilities, after the monopoles have proliferated, the final fate of the critical point is likely a state that spontaneously breaks the residual $SO(4)$ IR symmetry.

\section{Velocity Anisotropy in QED$_3$}
\label{app:aniso}
\begin{figure}[h]
\captionsetup{justification=raggedright}
\centering
\subcaptionbox{}
{\includegraphics[width=0.7\columnwidth]{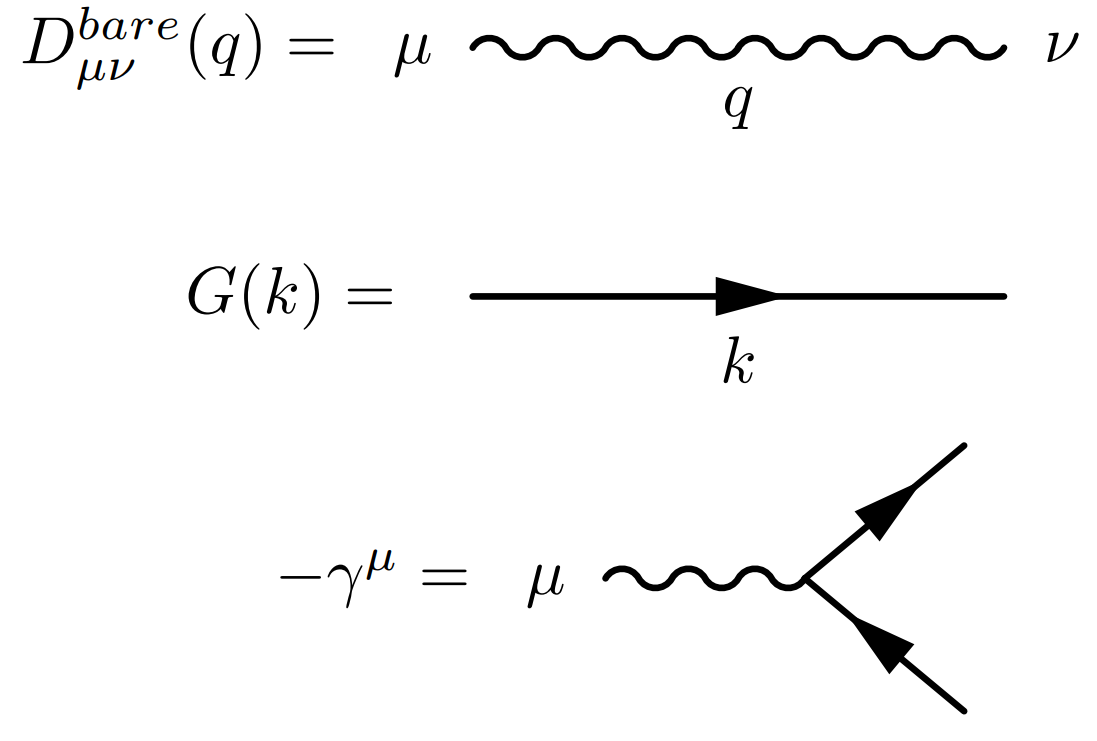}}
\subcaptionbox{}
{\includegraphics[width=0.7\columnwidth]{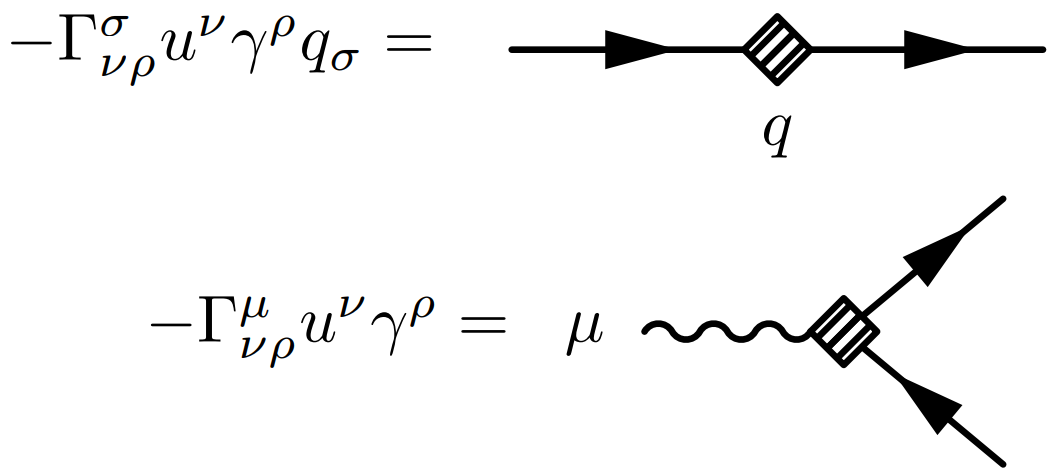}}
\caption{(a) Green's functions and photon-fermion vertex. (b) Velocity anisotropy corrections to the photon-fermion vertex.
}
\label{fig:greens}
\end{figure}

In this section, we will analyze the fate of velocity anisotropy in QED$_3$ under the renormalization group, in the large $N_f$ limit, similar to the analysis done in \cite{vafek_prb_2002,vafek_prl_2002,hermele_prb_2005,hermele_prb_2005_erratum}.
To begin, we have the Euclidean Lagrangian
\begin{equation}
    \mathcal{L}_E=\ol{\Psi}(-i\gamma^{\mu}(\del_{\mu}+ia_{\mu}))\Psi+\frac{1}{4e^2}f_{\mu\nu}^2,
\end{equation}
where $\gamma^{\mu}=(\tau^3,-\tau^2,\tau^1)$ and $\ol{\Psi}=i\Psi^{\dagger}\tau^3$. 
We consider the presence of a generic velocity anisotropy perturbation,
\begin{equation}
    \delta\mathcal{L}_E=\Gamma_{\nu\rho}^{\sigma}\ol{\Psi}\mu^{\nu}(-i\gamma^{\rho}(\del_{\sigma}+ia_{\sigma}))\Psi.
\end{equation}
The fermion propagator is given by
\begin{equation}
    G(k)=\frac{\slashed{k}}{k^2},
\end{equation}
and the bare photon propagator is
\begin{equation}
    D^{bare}_{\mu\nu}(q)=\frac{1}{q^2}\left(\delta_{\mu\nu}-\dfrac{q_{\mu}q_{\nu}}{q^2}\right)+\frac{16(\xi-1)}{N_f}\dfrac{q_{\mu}q_{\nu}}{q^2}+\mathcal{O}(q^2),    
\end{equation}
where we have introduced a nonlocal gauge-fixing parameter $\xi$.
The relevant diagrams are shown in Fig.~\ref{fig:greens}(a).

Furthermore, we also have the velocity anisotropy corrections to the fermion propagator, shown in Fig.~\ref{fig:greens}(b).

\begin{figure}[t]
\captionsetup{justification=raggedright}
\centering
\subcaptionbox{}
{\includegraphics[width=\columnwidth]{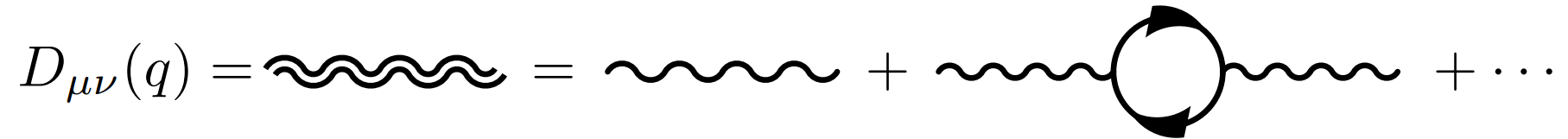}}
\subcaptionbox{}
{\includegraphics[width=0.6\columnwidth]{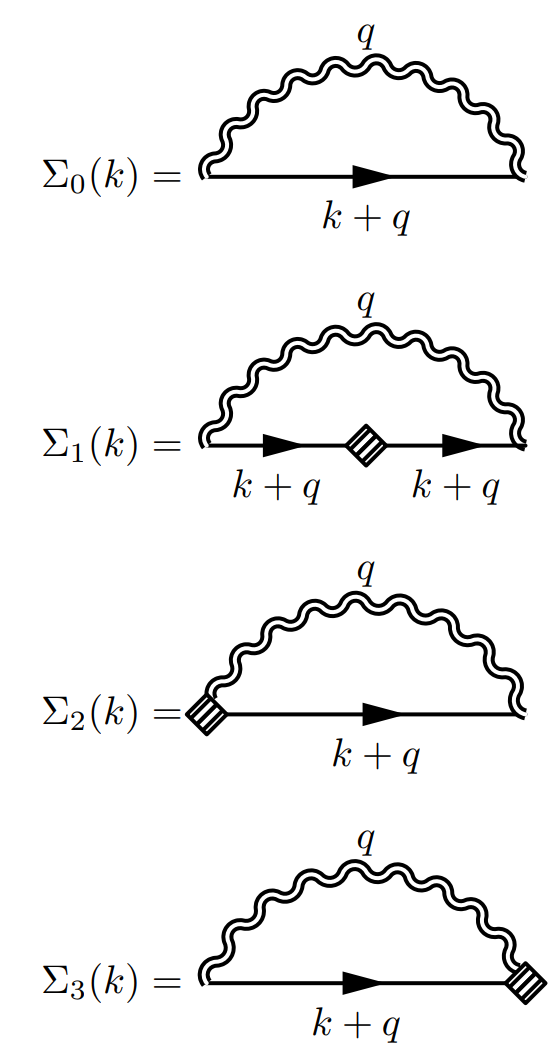}}
\caption{(a) Resummed effective gauge propagator and (b) self-energy corrections.
}
\label{fig:self_energ}
\end{figure}

In order to avoid resumming the same bubble diagrams each time, we will define an effective gauge propagator to leading order in $1/N_f$, obtained after resummation of the fermion bubbles as shown in Fig.~\ref{fig:self_energ}(a).

Extracting the leading contribution from the geometric series, we obtain
\begin{equation}
    D_{\mu\nu}(q)=\frac{16}{N_fq}\left(\delta_{\mu\nu}+\xi\dfrac{q_{\mu}q_{\nu}}{q^2}\right)+\mathcal{O}(q^2).
\end{equation}
Combining the above with the fermion Green's function and interaction vertices, the $1/N_f$ expansion for any correlator can be calculated.
In the following, we will suppress the flavor index $\nu$ in the vertices in Fig.~\ref{fig:greens}(b), with the understanding that to first-order in large $N_f$, there is no mixing between different flavor anisotropies $\mu^{\nu}(\cdots)$ corresponding to different $\nu$. 
Note that while in the main text, $\mu$ acts on the $2$ valley degrees of freedom, here we will generalize $\mu$ to act on any flavor degrees of freedom.
To find the flow of the $\Gamma$'s, we will need to calculate the self-energy corrections arising from the velocity anisotropy vertex, which is shown in Fig.~\ref{fig:self_energ}(b).
We will calculate the total self energy correction $\Sigma(k)=\Sigma_0(k)+\Sigma_1(k)+\Sigma_2(k)+\Sigma_3(k)$ to linear order in the velocity anisotropy $\Gamma$ and $1/N_f$.
To begin, we will first find the $1/N_f$ correction in $\Sigma_0$, given by (setting first the gauge fixing parameter $\xi$ to zero for simplicity)
\begin{align}
    \Sigma_0^{\xi=0}(k)&=\dfrac{16}{N_f}\int_{q}D_{\mu\nu}(q)\gamma^{\mu}G(k)\gamma^{\nu}\nonumber\\
    &=-\dfrac{16}{N_f}\int_q\dfrac{\slashed{k}+\slashed{q}}{q(k+q)^2}\nonumber\\
    &=-\dfrac{16}{N_f}\int_{q,x}\dfrac{1}{2\sqrt{1-x}}\dfrac{(1-x)\slashed{k}+\slashed{q}}{(q^2+x(1-x)k^2)^{3/2}}\nonumber
    \\&=\frac{8}{3\pi^2N_f}\slashed{k}\ln\frac{k}{\Lambda}.
\end{align}
We have regularized the integral with a UV cutoff $\Lambda$.
In the above and hereafter, we have used Feynman parameters 
\begin{equation}
    \frac{1}{A^aB^b}=\frac{\Gamma(a+b)}{\Gamma(a)\Gamma(b)}=\int_0^1 dx\frac{x^{a-1}(1-x)^{b-1}}{(B+x(A-B))^{a+b}}
\end{equation} taking $q^2=B$, $(k+q)^2=A$m and then shifting $q\rightarrow q-kx$. 
To find the gauge dependent part, we use that 
\begin{equation}
    \gamma^{\mu}\gamma^{\alpha}\gamma^{\nu}\frac{q_{\mu}q_{\nu}}{q^2}=2\slashed{q}\frac{q_{\alpha}}{q^2}-\gamma^{\alpha},
\end{equation} 
so for $\Sigma_0(k)\equiv \Sigma_0^{\xi=0}(k)+\Sigma_0^{\xi}(k)$, we calculate
\begin{align}
    \Sigma_0^{\xi}(k)&=\frac{16\xi}{N_f}\int_q\slashed{q}\dfrac{\slashed{k}+\slashed{q}}{(k+q)^2q^3}\slashed{q}\nonumber\\
    &=\frac{16\xi}{N_f}\int_q\dfrac{2\sla{q}q\cdot(k+q)}{(k+q)^2q^3}-\dfrac{\sla{k}+\sla{q}}{q(k+q)^2}.
\end{align}
The second term was calculated previously, while the first term can be calculated by setting \begin{equation}\int_q\frac{\sla{q}q\cdot(k+q)}{(k+q)^2q^3}=L(k)\frac{\sla{k}}{k^2},
\end{equation}
so that 
\begin{align}
    L(k)&=\int_q\dfrac{(k+q)\cdot q(k\cdot q)}{(k+q)^2q^3}\nonumber
    \\
    &=\int_{q,x}\dfrac{3\sqrt{1-x}}{2}\dfrac{(q+k(1-x))\cdot(q-kx)k\cdot(q-kx)}{(q^2+x(1-x)k^2)^{5/2}}\nonumber
    \\
    &=\int_{q,x}\dfrac{3\sqrt{1-x}}{2}\dfrac{\frac{1-5x}{3}k^2q^2-x^2(1-x)k^4}{(q^2+x(1-x)k^2)^{5/2}}\\
    &=\dfrac{k^2}{6\pi^2}\ln\dfrac{k}{\Lambda},
\end{align}
so the total correction is given by 
\begin{equation}
    \Sigma_0(k)=\frac{1}{\pi^2N_f}\left(\frac{8}{3}+8\xi\right)\sla{k}\ln\frac{k}{\Lambda}
\end{equation}
We will now calculate $\Sigma_{1,2,3}$.
To begin, we can find the corrections due to the couplings 
$\Gamma^{\sigma}_{\rho}\gamma^{\rho}\del_{\sigma}$
\begin{widetext}
\begin{align}
\Sigma_1(\Gamma^{\sigma}_{\rho},k)&=-\dfrac{16\Gamma^{\sigma}_{\rho}}{N_f}\int_q\dfrac{1}{q}\left(\gamma^{\mu}\dfrac{1}{\sla{k}+\sla{q}}(k+q)_{\sigma}\gamma^{\rho}\dfrac{1}{\sla{k}+\sla{q}}\gamma^{\mu}+\dfrac{\xi}{q^2}\sla{q}\dfrac{1}{\sla{k}+\sla{q}}(k+q)_{\sigma}\gamma^{\rho}\dfrac{1}{\sla{k}+\sla{q}}\sla{q}\right),
\\
\Sigma_2(\Gamma^{\sigma}_{\rho},k)&=
\dfrac{16\Gamma^{\sigma}_{\rho}}{N_f}\int_q\dfrac{1}{q}\left(\gamma^{\rho}\dfrac{1}{\sla{k}+\sla{q}}
\gamma^{\sigma}+\dfrac{\xi}{q^2}q_{\rho}\gamma^{\sigma}\dfrac{1}{\sla{k}+
\sla{q}}\sla{q}\right),
\\
\Sigma_3(\Gamma^{\sigma}_{\rho},k)&=
\dfrac{16\Gamma^{\sigma}_{\rho}}{N_f}\int_q\dfrac{1}{q}\left(\gamma^{\sigma}\dfrac{1}{\sla{k}+\sla{q}}
\gamma^{\rho}+\dfrac{\xi}{q^2}\sla{q}\dfrac{1}{\sla{k}+
\sla{q}}q_{\rho}\gamma^{\sigma}\right).
\end{align}
We will now consider each self energy term independently, performing the calculations when $\rho=\sigma$ and $\rho\ne \sigma$.
\subsection{$\Sigma_1(\Gamma^{\sigma}_{\rho},k)$}
Further expanding, we find
\begin{align}
    \Sigma_1(\Gamma^{\sigma}_{\rho},k)&=-\dfrac{16\Gamma^{\sigma}_{\rho}}{N_f}\int_q\dfrac{\gamma^{\mu}(\sla{k}+\sla{q})(k+q)_{\sigma}\gamma^{\rho}(\sla{k}+\sla{q})\gamma^{\mu}}{q(k+q)^4}+\dfrac{\sla{q}(\sla{k}+\sla{q})(k+q)_{\sigma}\gamma^{\rho}(\sla{k}+\sla{q})\sla{q}}{q^3(k+q)^4}.
\end{align}
Keeping only logarithmically divergent contributions, the first term can be expanded as
\begin{align}
        \Sigma_1^{\xi=0}(\Gamma^{\sigma}_{\rho},k)&=-\dfrac{16\Gamma^{\sigma}_{\rho}}{N_f}\int_{q,x}\dfrac{3x}{4\sqrt{1-x}}\dfrac{\gamma^{\mu}\gamma^{\lambda}\gamma^{\rho}\gamma^{\nu}\gamma^{\mu}(k+q)_{\lambda}(k+q)_{\sigma}(k+q)_{\nu}}{(q^2+x(k^2+2k\cdot q))^{5/2}}\nonumber
        \\
        &=-\dfrac{16\Gamma^{\sigma}_{\rho}}{N_f}\int_{q,x}\dfrac{3x\sqrt{1-x}}{4}\dfrac{\gamma^{\mu}\gamma^{\lambda}\gamma^{\rho}\gamma^{\nu}\gamma^{\mu}}{(q^2+x(1-x)k^2)^{5/2}}(k_{\lambda}q_{\sigma}q_{\nu}+q_{\lambda}k_{\sigma}q_{\nu}+q_{\lambda}q_{\sigma}k_{\nu}),
\end{align}
while the second term can be written as
\begin{align}
    \Sigma_1^{\xi}(\Gamma^{\sigma}_{\rho},k)&=-\dfrac{16\Gamma^{\sigma}_{\rho}}{N_f}\int_{q,x}\dfrac{15x\sqrt{1-x}}{4}\dfrac{(\sla{q}-\sla{k}x)(\sla{q}+\sla{k}(1-x))(q+k(1-x))_{\sigma}\gamma^{\rho}(\sla{q}+\sla{k}(1-x))(\sla{q}-\sla{k}x)}{(q^2+x(1-x)k^2)^{7/2}}
\end{align}
We will now focus on the cases in which $\rho=\sigma$ and when $\rho\ne \sigma$.
\subsubsection{$\Sigma_1(\Gamma^{\sigma}_{\rho},k)$ for $\rho=\sigma$}
After collecting all the nonzero contractions that yield logarithmic contributions, the gauge independent part ends up being
\begin{align}
     \Sigma_1^{\xi=0}(\Gamma^{\rho}_{\rho},k)&=-\dfrac{16\Gamma_{\rho}^{\rho}}{N_f}\int_{q,x}\dfrac{3x\sqrt{1-x}}{4}\dfrac{\frac{q^2}{3}(-2\sla{k}+k_\rho\gamma^{\rho})}{(q^2+x(1-x)k^2)^{5/2}}\nonumber\\
     &=\dfrac{8\Gamma_{\rho}^{\rho}}{15\pi^2N_f}(-2\sla{k}+k_{\rho}\gamma^{\rho})\ln\dfrac{k}{\Lambda}.
\end{align}
Collecting all the contributions for the gauge dependent part, $\Sigma_1^{\xi}(\Gamma^{\rho}_{\rho},k)$, yields
\begin{align}
    \Sigma_1^{\xi}(\Gamma^{\rho}_{\rho},k)&=-\dfrac{16\Gamma^{\sigma}_{\rho}}{N_f}\int_{q,x}\dfrac{15x\sqrt{1-x}}{4}q^4\dfrac{\frac{5-7x}{3}
    k_{\rho}\gamma^{\rho}-\sum_{\nu\ne \rho}\frac{2}{3}k_\nu\gamma^{\nu}
    }{(q^2+x(1-x)k^2)^{7/2}}\nonumber
    \\
&=\Gamma^{\rho}_{\rho}\left(\dfrac{8}{3\pi^2N_f}k_{\rho}\gamma^{\rho}-\dfrac{16}{3\pi^2N_f}\sum_{\nu\ne \rho}k_{\nu}\gamma^{\nu}\right)\ln\dfrac{k}{\Lambda}
\end{align}
\subsubsection{$\Sigma_1(\Gamma^{\sigma}_{\rho},k)$ for $\rho\ne\sigma$}
The gauge independent part can be calculated to be
\begin{align}
     \Sigma_1^{\xi=0}(\Gamma^{\sigma}_{\rho},k)&=-\dfrac{16\Gamma^{\sigma}_{\rho}}{N_f}\int_{q,x}\dfrac{3x\sqrt{1-x}}{4}\dfrac{3k_{\sigma}\gamma^{\rho}-2k_{\rho}\gamma^{\sigma}}{(q^2+x(1-x)k^2)^{5/2}}\nonumber
     \\
     &=\dfrac{8\Gamma_{\rho}^{\rho}}{15\pi^2N_f}(3k_{\sigma}\gamma^{\rho}-2k_{\rho}\gamma^{\sigma})\ln\dfrac{k}{\Lambda}.
\end{align}
The gauge dependent part $\Sigma_1^{\xi}(\Gamma^{\sigma}_{\rho},k)$ yields
\begin{align}
\Sigma_1^{\xi}(\Gamma^{\sigma}_{\rho},k)&=-\dfrac{16\Gamma^{\sigma}_{\rho}}{N_f}\int_{q,x}\dfrac{15x\sqrt{1-x}}{4}q^4
\dfrac{\frac{5-7x}{3}k_{\sigma}\gamma^{\rho}+\frac{2}{3}k_{\rho}\gamma^{\sigma} }
{(q^2+x(1-x)k^2)^{7/2}}\nonumber
\\
&=\Gamma^{\sigma}_{\rho}\left(\dfrac{8}{3\pi^2N_f}k_{\sigma}\gamma^{\rho}+\dfrac{16}{3\pi^2N_f}k_{\rho}\gamma^{\sigma}\right)\ln\dfrac{k}{\Lambda}.
\end{align}
\subsection{$\Sigma_{2,3}(\Gamma^{\sigma}_{\rho},k)$}
Further expanding $\Sigma_2$, we find
\begin{align}
    \Sigma_2(\Gamma^{\sigma}_{\rho},k)&=\dfrac{16\Gamma^{\sigma}_{\rho}}{N_f}\int_{q,x}\dfrac{1}{2\sqrt{1-x}}
    \dfrac{\gamma^{\rho}(\sla{q}+\sla{k}(1-x))\gamma^{\sigma}}{(q^2+x(1-x)k^2)^{3/2}}+\dfrac{3\sqrt{1-x}}{2}\dfrac{(q-kx)_{\sigma}\gamma^{\rho}(\sla{q}+\sla{k}(1-x))(\sla{q}-\sla{k})}{(q^2+x(1-x)k^2)^{-5/2}}.
\end{align}
$\Sigma_3$ is similar, except with the interaction insertion at a different point.
\begin{align}
    \Sigma_3(\Gamma^{\sigma}_{\rho},k)&=\dfrac{16\Gamma^{\sigma}_{\rho}}{N_f}\int_{q,x}\dfrac{1}{2\sqrt{1-x}}
    \dfrac{\gamma^{\sigma}(\sla{q}+\sla{k}(1-x))\gamma^{\rho}}{(q^2+x(1-x)k^2)^{3/2}}+\dfrac{3\sqrt{1-x}}{2}\dfrac{(\sla{q}-\sla{k})(\sla{q}+\sla{k}(1-x))(q-kx)_{\sigma}\gamma^{\rho}}{(q^2+x(1-x)k^2)^{-5/2}}.
\end{align}
\subsubsection{$\Sigma_2(\Gamma^{\sigma}_{\rho},k)$ for $\rho=\sigma$}
The gauge independent part ends up being
\begin{align}
     \Sigma_2^{\xi=0}(\Gamma^{\rho}_{\rho},k)&=\dfrac{16\Gamma^{\sigma}_{\rho}}{N_f}\int_{q,x}\frac{\sqrt{1-x}}{2}\dfrac{k_{\rho}\gamma^{\rho}-\sum_{\nu\ne\rho}k_{\nu}\gamma^{\nu}}{(q^2+x(1-x)k^2)^{3/2}}\nonumber
     \\
     &=-\Gamma_{\rho}^{\rho}\dfrac{8}{3\pi^2N_f}
     \left(k_{\rho}\gamma^{\rho}-\sum_{\nu\ne\rho}k_{\nu}\gamma^{\nu}\right)\ln\frac{k}{\Lambda},
\end{align}
while the gauge dependent part is
\begin{align}
    \Sigma_2^{\xi}(\Gamma^{\rho}_{\rho},k)&=\dfrac{16\Gamma^{\sigma}_{\rho}}{N_f}\int_{q,x}\dfrac{3\sqrt{1-x}}{2}\dfrac{
    \frac{1-5x}{3}k_{\rho}\gamma^{\rho}-\sum_{\nu\ne\rho}\frac{1}{3}k_{\nu}\gamma^{\nu}}{(q^2+x(1-x)k^2)^{-5/2}}\nonumber
    \\&=\Gamma^{\rho}_{\rho}\dfrac{8}{3\pi^2N_f}\sla{k}\ln\dfrac{k}{\Lambda}.
\end{align}
Note for $\rho=\sigma$, $\Sigma_2=\Sigma_3$.
\subsubsection{$\Sigma_2(\Gamma^{\sigma}_{\rho},k)$ for $\rho\ne\sigma$}
We can find the gauge independent part,
\begin{align}
     \Sigma_2^{\xi=0}(\Gamma^{\sigma}_{\rho},k)&=\dfrac{16\Gamma^{\sigma}_{\rho}}{N_f}\int_{q,x}\frac{\sqrt{1-x}}{2}\dfrac{k_{\sigma}\gamma^{\rho}+k_{\rho}\gamma^{\sigma}+ik_{\nu\ne\rho,\sigma}\mathbb{I}}{(q^2+x(1-x)k^2)^{3/2}}=-\Gamma_{\rho}^{\sigma}\dfrac{8}{3\pi^2N_f}
     \left(k_{\sigma}\gamma^{\rho}+k_{\rho}\gamma^{\sigma}+ik_{\nu\ne\rho,\sigma}\mathbb{I}\right)\ln\frac{k}{\Lambda},
\end{align}
where $\Sigma_3^{\xi=0}$ is identical, except with an opposite sign for $ik_{\nu\ne\rho,\sigma}\mathbb{I}$,
while the gauge dependent part is
\begin{align}
    \Sigma_2^{\xi}(\Gamma^{\sigma}_{\rho},k)&=\dfrac{16\Gamma^{\sigma}_{\rho}}{N_f}\int_{q,x}\dfrac{3\sqrt{1-x}}{2}\dfrac{
    \frac{1-5x}{3}k_{\sigma}\gamma^{\rho}+\frac{1}{3}k_{\rho}\gamma^{\sigma}-\frac{1}{3}ik_{\nu\ne\rho,\sigma}\mathbb{I}}{(q^2+x(1-x)k^2)^{-5/2}}
=\Gamma^{\sigma}_{\rho}\dfrac{16}{3\pi^2N_f}\left(k_{\sigma}\gamma^{\rho}-k_{\rho}\gamma^{\sigma}-ik_{\nu\ne\rho,\sigma}\mathbb{I}\right)\ln\dfrac{k}{\Lambda}.
\end{align}
As with $\Sigma_2^{\xi=0}$, $\Sigma_3^{\xi}$ is equal to $\Sigma_2^{\xi}$, except with an opposite sign for $ik_{\nu\ne\rho,\sigma}\mathbb{I}$.
\subsection{Renormalization}
Collecting the above results and restoring the valley index $(\nu)$ in $\Gamma^{\sigma}_{\nu\rho}$, we obtain the vertex function
\begin{align}
    \Gamma=&\left(1-\dfrac{8+24\xi}{3\pi^2N_f}\ln\frac{k}{\Lambda}\right)\sla{k}+\sum_{\nu}\mu^{\nu}\sum_{\rho\ne\sigma}\Gamma^{\sigma}_{\nu\rho}k_{\sigma}\gamma^{\rho}+\left(\Gamma^{\sigma}_{\nu\rho}\dfrac{56-120\xi}{15\pi^2N_f}+\Gamma_{\rho}^{\nu\sigma}\dfrac{32}{5\pi^2N_f}\right)k_{\sigma}\gamma^{\rho}\ln\frac{k}{\Lambda}\nonumber\\
    &\sum_{\nu}\mu^{\nu}\sum_{\rho}\Gamma_{\rho}^{\nu\rho}k_{\rho}\gamma^{\rho}+\left(\Gamma^{\rho}_{\nu\rho}\dfrac{88-120\xi}{15\pi^2N_f}-\sum_{\lambda\ne\rho}\Gamma_{\nu\lambda}^{\lambda}\dfrac{64}{15\pi^2N_f}\right)k_{\rho}\gamma^{\rho}\ln\frac{k}{\Lambda}
\end{align}

\end{widetext}

The renormalization two point vertex is related to the above bare through a field renormalization $Z_{\Psi}$, such that $\Gamma_R=Z_{\Psi}\Gamma$.
For our renormalization condition, we demand that at a momentum scale $p$, $\Gamma_R(p)$ has the form $\sla{p}+\sum_{\nu}\mu^{\nu}\sum_{\rho,\sigma}(\Gamma_{\nu\rho}^{\sigma})_Rk_{\sigma}\gamma^{\rho}$, which gives us 
\begin{equation}
Z_{\Psi}=1+\frac{8+24\xi}{3\pi^2N_f}\ln\frac{p}{\Lambda},
\end{equation} from which we can derive the renormalized velocities:
\begin{widetext}
\begin{itemize}
    \item For $\rho=\sigma$,
    \begin{equation}
        (\Gamma_{\nu\rho}^{\rho})_R(p)=\Gamma_{\nu\rho}^{\rho}\left(1+\frac{8+24\xi}{3\pi^2N_f}\ln\frac{p}{\Lambda}\right)+\left(\Gamma^{\rho}_{\nu\rho}\dfrac{88-120\xi}{15\pi^2N_f}-\sum_{\lambda\ne\rho}\Gamma_{\nu\lambda}^{\lambda}\dfrac{64}{15\pi^2N_f}\right)\ln\frac{p}{\Lambda},
    \end{equation}
    from which allowing $p\sim e^{-l}$ yields the beta function 
    \begin{equation}
        \dfrac{d}{dl}(\Gamma_{\nu\rho}^{\rho})_R=-\frac{128}{15\pi^2N_f}\Gamma_{\nu\rho}^{\rho}+\frac{64}{15\pi^2N_f}\sum_{\lambda\ne\rho}\Gamma_{\nu\lambda}^{\lambda}.
    \end{equation}
\item For $\rho\ne\sigma$,
    \begin{equation}
        (\Gamma_{\nu\sigma}^{\rho})_R(p)=\Gamma_{\nu\rho}^{\sigma}\left(1+\frac{8+24\xi}{3\pi^2N_f}\ln\frac{p}{\Lambda}\right)+
        \left(\Gamma^{\sigma}_{\nu\rho}\dfrac{56-120\xi}{15\pi^2N_f}+\Gamma_{\rho}^{\nu\sigma}\dfrac{32}{5\pi^2N_f}\right)\ln\frac{p}{\Lambda},
    \end{equation}
    which gives
    \begin{equation}
        \dfrac{d}{dl}(\Gamma_{\nu\sigma}^{\rho})_R=-\frac{32}{5\pi^2N_f}\Gamma_{\nu\sigma}^{\rho}-\frac{32}{5\pi^2N_f}\Gamma_{\nu\rho}^{\sigma}.
    \end{equation}
\end{itemize}
\end{widetext}
Therefore, we see that all perturbations of the form $\dfrac{d}{dl}(\Gamma_{\nu\rho}^{\rho})_R$ are irrelevant.
Under the above renormalization scheme, the first bullet point yields the flow eigenvectors described by couplings $-i\Gamma_1\ol{\Psi}\mu(\gamma^1D_1-\gamma^2D_2)\psi$ with \begin{equation}
    \beta_{\Gamma_1}=-\frac{64}{5\pi^2N_f}\Gamma_1,
\end{equation} and 
$-i\Gamma_2\ol{\Psi}\mu(\gamma^{\mu}D_{\mu})\Psi$, which is marginal to one loop in $\frac{1}{N_f}$.
The second bullet point gives us the couplings $-i\Gamma_3\ol{\Psi}\mu(\gamma^1D_2+\gamma^2D_1)\Psi$ with \begin{equation}
\beta_{\Gamma_3}=-\frac{64}{5\pi^2N_f}\Gamma_3,    
\end{equation}
and $-i\Gamma_4\ol{\Psi}\mu(\gamma^1D_2-\gamma^2D_1)\Psi$, which is marginal. 
Furthermore, as desired, our renormalization group analysis is fully gauge invariant as there is no dependence on the gauge fixing parameter $\xi$.

The above suggests that the QED$_3$ fixed point is not destabilized by velocity anisotropies, 
except for the two marginal anisotropic perturbations above.
However, we will demonstrate that the two marginal operators $\Gamma_2$ and $\Gamma_4$ are redundant operators, and therefore, do not affect the conformal fixed point.
In more detail, a redundant operator is any operator in the action that can be removed by a continuous definition of the fields \cite{weinberg_textbook_qft},
\begin{equation}\label{eq:redun}
 \Psi^l(x)\rightarrow \Psi^l(x)+\epsilon F^l(\Psi(x),\del_{\mu}\Psi(x),\cdots).
\end{equation}
Such operators do not affect any observables of the theory.
Furthermore, the field redefinition in Eq.~\eqref{eq:redun} changes the Lagrangian as 
\begin{equation}
    \delta \mathcal{L}=\epsilon\sum_l\dfrac{\delta\mathcal{L}}{\delta\Psi_l}F^l(\Psi(x),\del_{\mu}\Psi(x),\cdots).
\end{equation}
Thus, a redundant operator is equivalently one that vanishes on the equations of motion, $\frac{\delta\mathcal{L}}{\delta\Psi}=0$.
Strictly speaking, the renormalization group behavior of a redundant operator is scheme dependent and not universal \cite{wegner_1974}.
In our renormalization group flow, both $\Gamma_2$ and $\Gamma_4$ are marginal as they arise from a constant rescaling (or reparametrization) of $\Psi$ at the fixed point. 
Specifically, $\Gamma_2$ can be eliminated from the action at the fixed point theory by redefining $\Psi\rightarrow {\Psi}-\frac{\Gamma_2}{2}\mu{\Psi}$, followed by a field renormalization, while $\Gamma_4$ can be eliminated by redefining ${\Psi}\rightarrow\Psi+i\frac{\Gamma_4}{2}\mu\gamma^0\Psi$ and then rescaling the time coordinate.
\section{LSM Anomaly Matching}
\label{app:lsm}
We follow the approach outlined in \cite{ye_scipost_2022}.
Recall that from
Eq.~\eqref{eq:anom_match} in the main text
we must match the anomalies as elements of the group cohomology $H^4(G_{UV},U(1)_T)$.
The $U(1)$ coefficients complicate the process, but for the groups we are considering, $H^4(G_{UV},U(1)_T)$ is just a product of $\mathbb{Z}_2$ factors.
Then, it is much simpler to calculate the IR and UV anomalies as elements of $H^4(G_{UV},\mathbb{Z}_2)$.
This equates to viewing
\begin{equation}
    \eta[G_{UV/IR}]=\exp\left(i\pi \Omega[G_{UV/IR}]\right),
\end{equation}
where $\Omega[G_{UV/IR}]\in H^4(G_{UV/IR},\mathbb{Z}_2)$.
The operation $\exp(i \pi \bullet)$ can be considered as an inclusion $\mathbb{Z}_2\rightarrow U(1)$.

However, in practice, we will use an analogous relation to Eq.~\eqref{eq:anom_match} but instead consider the UV anomaly and the pullback of the IR anomaly as elements of $H^5(G_{UV},\mathbb{Z}_2)$. 
Though this matching occurs in one higher dimension, it is much simpler than confirming Eq.~\eqref{eq:anom_match}.
The reason is that the induced map on cohomology from $\exp(i\pi\bullet)$ (for which we will use the same notation),
\begin{equation}
    {\exp(i\pi\bullet)}\;:\;H^4(G_{UV},\mathbb{Z}_2)\rightarrow H^4(G_{UV},U(1)_{T})
\end{equation} is not injective, so that confirming the anomaly matching and pullback condition in $H^4(G_{UV},\mathbb{Z}_2)$ is not equivalent to showing that $\eta[G_{UV}]=\varphi^*\eta[G_{IR}]\in H^4(G_{UV},U(1)_{T})$. 
The anomaly matching in  $H^4(G_{UV},\mathbb{Z}_2)$ is a sufficient but not necessary condition.
In other words, one can have
\begin{equation}
      (\eta[G_{UV}])=\varphi^*(\eta[G_{IR}])
\end{equation}
but 
\begin{equation}
        \Omega[G_{UV}]\ne\varphi^*(\Omega[G_{IR}]).
\end{equation}
The correct matching condition that is equivalent to Eq.~\eqref{eq:anom_match} is 
\begin{equation}
    \label{eq:anom_match_sq1}
    \mathcal{SQ}^1(\Omega[G_{UV}])=\varphi^*\mathcal{SQ}^1(\Omega[G_{IR}]).
\end{equation}
where the operation $\mathcal{SQ}^1$ is a composition of maps $\mathcal{SQ}^1=\rho_2\circ\beta\circ{\exp(i\pi\bullet)}$.
We have defined  $\rho_2$ to be reduction $\mathbb{Z}\rightarrow \mathbb{Z}_2$, $\beta$ as the Bockstein homomorphism taking $H^n(G,U(1)_T)\rightarrow H^{n+1}(G,\mathbb{Z})$, and ${\exp(i\pi\bullet)}$ as the map in cohomology,
\begin{equation}
{\exp(i\pi\bullet)}\;:\;H^4(G_{UV},\mathbb{Z}_2)\rightarrow H^4(G_{UV},U(1)_{T}).
\end{equation}
In particular $\exp(i\pi\bullet)$ maps the $\mathbb{Z}_2$ valued class $\Omega[G]$ into the $U(1)$ valued anomaly $\eta[G]$. More details can be found in \cite{ye_scipost_2022}.

\subsection{The form of $\eta$ and $\Omega$ for $G_{UV/IR}$}
To begin, we note that in all of our UV models, there is no lattice LSMOH theorem that applies, as we have trivial bosons, which are integer $SO(3)$ spin and Kramers singlets, occupying each unit cell and high symmetry point.
Note in the kagome atomic insulator, the UV symmetry is no longer $SO(3)\times \mathbb{Z}_2^T\times G_{lattice}$ but $SO(2)\times\mathbb{Z}_2^T\times G_{lattice}$ due to the spin polarization of the vacuum in our parton description. 
However, in all cases, we have that $\eta[G_{UV}]$ and $\Omega[G_{UV}]$ are trivial.

Now in the IR, we have $N_f=4$ QED$_3$ with $G_{IR}={SO(6)\times U(1)_{top}}/{\mathbb{Z}_2}$.
However, there is not yet a known form of the anomaly $G_{IR}$ in terms of characteristic classes
for the $G_{IR}$ anomaly for $N_f=4$ QED$_3$.
Instead, we will consider a slightly simplified description of the anomaly by coupling $SO(6)$ and $U(1)_{top}=SO(2)$ gauge fields, considering the QED$_3$ CFT as a Stiefel liquid \cite{zou_prx_2021}, which provides a dual description of the theory in terms of a nonlinear sigma model.
Allowing time reversal, we must then consider an $O(6)_T\times O(2)_T$ gauge bundle, where the subscript $T$ indicates that improper rotations in $O(2)$ or $O(6)$ must be composed with a spacetime orientation reversal symmetry.
Mathematically, this constrains
\begin{equation}
    w_1^{TM}+w_1^{O(6)}+w_1^{O(2)}=0\pmod{2},
\end{equation}
in terms of the Stiefel-Whitney classes of the tangent, $O(6)$, and $O(2)$ bundles. 
The 't Hooft anomaly is then characterized by an SPT bulk, \cite{zou_prx_2021}
\begin{widetext}
\begin{align}
\exp\left[i\pi\left(w_4^{O(6)}+w_2^{O(6)}\left(w_2^{O(2)}+(w_1^{O(2)})^2\right)+\left((w_2^{O(2)})^2+w_2^{O(2)}(w_1^{O(2)})^2+(w_1^{O(2)})^4\right)\right)\right].    
\end{align}
\end{widetext}
Note that the above symmetry group $O(6)_T\times O(2)_T$ does not consider the discrete $\mathbb{Z}_2$ quotient and is therefore larger than the faithful IR symmetry group ${O(6)_T\times O(2)_T}/{\mathbb{Z}_2}$.
Consequently, the above anomaly is a coarser description of the actual anomaly of $N_f=4$ QED$_3$.
In a different physical system, the above anomaly could still be accurate if there are additional gapped, trivial vector degrees of freedom in the IR.
In any case, we will use the enlarged symmetry as only then is there a (known) expression for the anomaly that allows us to examine the anomaly matching explicitly.
Using the englarged $\tilde{G}_{IR}=O(6)_T\times O(2)_T$, we have
\begin{widetext}
\begin{align}
\eta[\tilde{G}_{IR}]&=\exp\left[i\pi\left(w_4^{O(6)}+w_2^{O(6)}\left(w_2^{O(2)}+(w_1^{O(2)})^2\right)+\left((w_2^{O(2)})^2+w_2^{O(2)}(w_1^{O(2)})^2+(w_1^{O(2)})^4\right)\right)\right]\nonumber\\
\implies \Omega[\tilde{G}_{IR}]&=w_4^{O(6)}+w_2^{O(6)}\left(w_2^{O(2)}+(w_1^{O(2)})^2\right)+\left((w_2^{O(2)})^2+w_2^{O(2)}(w_1^{O(2)})^2+(w_1^{O(2)})^4\right).
\end{align}
\end{widetext}
For here onwards and the calculation of the pullback, we will drop the tilde and consider $G_{IR}=O(6)_T\times O(2)_T$ for simplicity.
Mathematically, this is equivalent to taking the UV to IR embedding $\varphi$, factorizing into two parts
\begin{equation}
    \varphi\;:\;G_{UV}\rightarrow O(6)_T\times O(2)_T\rightarrow \frac{ O(6)_T\times O(2)_T}{\mathbb{Z}_2},
\end{equation} and considering the pullback of only the first part of the embedding.

Lastly, we note that under the operator $\mathcal{SQ}^1$, the IR anomaly takes the form
\begin{widetext}
\begin{align}
\mathcal{SQ}^1(\Omega[{G}_{IR}])=  w_5^{O(6)} + w_4^{O(6)}w_1^{O(2)} + w_3^{O(6)}\left(w_2^{O(2)} + (w_1^{O(2)})^2\right) + 
&w_2^{O(6)}(w_1^{O(2)})^3 +\left((w_2^{O(2)})^2 w_1^{O(2)} + (w_1^{O(2)})^5\right)+ \nonumber\\
&w_1^{O(6)}\left((w_2^{O(2)})^2 + w_2^{O(2)}(w_1^{O(2)})^2 + (w_1^{O(2)})^4\right).
\label{eq:anom_sq1}
\end{align}
\end{widetext}
Then, Eq.~\eqref{eq:anom_match_sq1} amounts to checking
\begin{equation}
    \varphi^*\mathcal{SQ}^1(\Omega[{G}_{IR}])=0
\end{equation} for the three critical points considered in our paper.
\subsection{Calculating the Pullbacks}
We begin by performing anomaly matching for the square lattice critical point.
We note for the square lattice (and all the cases we consider), it is clear from Table~\ref{table:monopole_fund_bbh} that the symmetry $O(6)_T=O(3)_v\times O(3)_f$ factorizes into $O(3)_{v,f}$ blocks for the valley and pseudospin symmetries acting on the monopole operators.
Repeatedly applying the Whitney product formula allows us to calculate the pullbacks of each block, in terms of cohomology classes of $G_{UV}=p4m\times O(3)_T$
\begin{align}
    \varphi^*(w_1^{O(3)_v})&=0\\
    \varphi^*(w_2^{O(3)_v})&=r^2\\
    \varphi^*(w_3^{O(3)_v})&=0\\
    \varphi^*(w_1^{O(3)_f})&=t\\
    \varphi^*(w_2^{O(3)_f})&=w_2^{O(3)_T}\\
    \varphi^*(w_2^{O(3)_f})&=w_3^{O(3)_T}.
\end{align}
We have defined $t\in H^1(\mathbb{Z}_2^T,\mathbb{Z}_2)$ as the gauge
field corresponding time-reversal symmetry and $r\in H^1(p4m,\mathbb{Z}_2)$ as the reflection gauge field.
Note that $O(3)_T$ represents the $SO(3)\times \mathbb{Z}_2^T$ transformation.
As $\mathbb{Z}_2^T$ acts as minus one in the $SO(3)$ block, we have $w_3^{O(3)_T}=w_3^{SO(3)}+tw_2^{SO(3)}+t^3$.
Assembling the above, one can find
\begin{align}
    \varphi^*(w_1^{O(6)})&=t,\\
    \varphi^*(w_2^{O(6)})&=r^2+w_2^{O(3)_T},\\
    \varphi^*(w_3^{O(6)})&=tr^2+w_3^{O(3)_T},\\
    \varphi^*(w_4^{O(6)})&=r^2w_2^{O(3)_T},\\
    \varphi^*(w_5^{O(6)})&=r^2w_3^{O(3)_T}.
\end{align}
    Moreover, we can find the $O(2)$ contributions
\begin{equation}
        \varphi^*(w_1^{O(2)})=r,\quad
    \varphi^*(w_2^{O(2)})=0.
\end{equation}
Combining the above and substituting into Eq.~\eqref{eq:anom_sq1}, we indeed obtain Eq.~\eqref{eq:anom_match_sq1}, $\varphi^*\mathcal{SQ}^1(\Omega[{G}_{IR}])=0$.

For the breathing honeycomb lattice, we introduce an additional gauge field $c$ for $C_2$ rotation, from which we find from Table~\ref{table:monopole_fund_c6}
\begin{align}
    \varphi^*(w_1^{O(3)_v})&=0\\
    \varphi^*(w_2^{O(3)_v})&=r^2+c^2+rc\\
    \varphi^*(w_3^{O(3)_v})&=cr(r+c)\\
    \varphi^*(w_1^{O(3)_f})&=t\\
    \varphi^*(w_2^{O(3)_f})&=w_2^{O(3)_T}\\
    \varphi^*(w_2^{O(3)_f})&=w_3^{O(3)_T}.
\end{align}
The $O(2)$ contributions are
\begin{equation}
    \varphi^*(w_1^{O(2)})=r,\quad
    \varphi^*(w_2^{O(2)})=c^2+cr,
\end{equation}
from which we can confirm $\varphi^*\mathcal{SQ}^1(\Omega[{G}_{IR}])=0$.

For the breathing kagome lattice, we will use the transformations from Table~\ref{table:monopole_fund_c3},
\begin{align}
    \varphi^*(w_1^{O(3)_v})&=0\\
    \varphi^*(w_2^{O(3)_v})&=r^2\\
    \varphi^*(w_3^{O(3)_v})&=0\\
    \varphi^*(w_1^{O(3)_f})&=0\\
    \varphi^*(w_2^{O(3)_f})&=w_2^{O(2)_T}+t^2\\
    \varphi^*
    (w_2^{O(3)_f})&=tw_2^{O(2)_T},
\end{align}
where here, $O(2)_T$ refers to the UV spin rotation symmetry that acts on the monopoles $\phi_{4,5}$.
Note that from the embedding of the UV symmetries, we have $w_2^{O(2)_T}=w_2^{SO(2)}+r^2+rt$.
The $O(2)$ contributions are
\begin{equation}
    \varphi^*(w_1^{O(2)})=r+t,\quad
    \varphi^*(w_2^{O(2)})=0,
\end{equation}
from which we can find \begin{equation}
    \label{eq:kag_sq1}
    \varphi^*\mathcal{SQ}^1(\Omega[{G}_{IR}])=w_2^{O(2)_T}rt^2+r^2t^3+t^3r^2
\end{equation}
We can simplify the above using the Steenrod square as the first term vanishes,
\begin{align}
    w_2^{O(2)_T}rt^2=Sq^1(w_2^{O(2)_T}t^2)+w_2^{O(2)_T}t^3=0,
\end{align}
as for an $O(2)$ bundle, $Sq^1(w_{2})=w_1w_{2}$.
The second and third terms on the RHS of Eq.~\eqref{eq:kag_sq1} also vanish, as
\begin{equation}
    r^2t^3+r^3t^2=Sq^1(r^2t^2)+r^3t^2+r^3t^2=0.
\end{equation}
Therefore, as desired, all of the critical points we have considered are consistent with anomaly matching.
\bibliography{sources}
\end{document}